\documentclass[%
 aip,
 amsmath,amssymb,
 reprint,%
]{revtex4-1}

\usepackage{graphicx}% Include figure files
\usepackage{dcolumn}% Align table columns on decimal point
\usepackage{bm}% bold math
\usepackage[utf8]{inputenc}
\usepackage[T1]{fontenc}
\usepackage{mathptmx}
\usepackage{etoolbox}

\usepackage{caption}
\usepackage{subcaption}
\usepackage{tabularray}
\usepackage{xcolor}

\makeatletter
\def\@email#1#2{%
 \endgroup
 \patchcmd{\titleblock@produce}
  {\frontmatter@RRAPformat}
  {\frontmatter@RRAPformat{\produce@RRAP{*#1\href{mailto:#2}{#2}}}\frontmatter@RRAPformat}
  {}{}
}%
\makeatother
\begin{document}

\preprint{AIP/123-QED}

\title[Damping of three-dimensional waves on coating films dragged by moving substrates]{Damping of three-dimensional waves on coating films dragged by moving substrates}
% Force line breaks with \\
\author{David Barreiro-Villaverde}
\email{david.barreiro1@udc.es}
\affiliation{Universidade da Coruña, CITIC Research, Campus de Elviña, 15071 A Coruña, Spain}
\affiliation{von Karman Institute for Fluid Dynamics, B-1640 Waterloosesteenweg 72, Sint-Genesius-Rode, Belgium}

\author{Anne Gosset}
\affiliation{Universidade da Coruña, Campus Industrial de Ferrol, CITENI, 15403 Ferrol, Spain}%

\author{Marcos Lema}
\affiliation{Universidade da Coruña, Campus Industrial de Ferrol, CITENI, 15403 Ferrol, Spain}%

\author{Miguel A. Mendez}
\affiliation{von Karman Institute for Fluid Dynamics, B-1640 Waterloosesteenweg 72, Sint-Genesius-Rode, Belgium}

\date{\today}  % It is always \today, today,
               %  but any date may be explicitly specified

\begin{abstract}

%\textcolor{black}{coatings} and paints often feature interfacial defects due to disturbances during the deposition process which, if they persist until solidification, worsen the product quality. In this article, we investigate the \textcolor{black}{stability of} a thin liquid film dragged by a vertical moving substrate against gravity, a flow configuration found in a variety of coating processes. The receptivity of the liquid film to three-dimensional disturbances is discussed with Direct Numerical Simulations (DNS), an in-house non-linear Integral Boundary Layer (IBL) film model, and Linear Stability Analysis (LSA). The thin film model, which is successfully validated with the DNS computations, implements a pseudo-spectral approach for the capillary terms that allows for investigating non-periodic capillarity dominated flows. The combination of these numerical tools allows for describing the mechanisms of capillary and non-linear damping, as well as for identifying the instability threshold o of the coating processes. Transverse modulations are found to be beneficial for the damping of two-dimensional waves within the range of operational parameters conditions considered in this study, \textcolor{black}{(and)} typical of \textcolor{black}{(the)} air-knife and slot-die coating processes.

\textcolor{black}{Paints and coatings} often feature interfacial defects due to disturbances during the deposition process which, if they persist until solidification, \textcolor{black}{worsens} the product quality. In this article, we investigate the stability of a thin liquid film dragged by a vertical substrate moving against gravity, a fundamental flow configuration in various coating processes. The receptivity of the liquid film to three-dimensional disturbances is analyzed with Direct Numerical Simulations (DNS) and an in-house Integral Boundary Layer (IBL) film model. The latter was used for Linear Stability Analysis (LSA) and nonlinear wave propagation analysis. The numerical implementation of the IBL film model combines a finite volume formulation with a pseudo-spectral approach for the capillary terms that allows for investigating non-periodic surface tension-dominated flows. Both the model and the numerical solver were successfully validated with DNS computations. The combination of these numerical tools allows for describing the mechanisms of capillary and nonlinear damping and identifying the instability threshold of the coating processes. The results show that transverse modulations can be beneficial for damping two-dimensional waves within the range of operational conditions considered in this study, which are relevant to air-knife and slot-die coating.

\end{abstract}

\maketitle

%% main text
\section{Introduction}
\label{sec:intro}

Liquid films dragged by moving substrates are often found in industrial coating and painting processes. In these processes, thickness inhomogeneities reduce the quality and performance of the final products and are thus considered defects. \textcolor{black}{Therefore, the damping of non-uniformities soon after liquid deposition and before solidification is fundamental.}

Liquid film flows are generally unstable and naturally develop interfacial waves that evolve in time and space, even at very low Reynolds numbers. The analysis of thin films instabilities began in the \textcolor{black}{1910s} with the pioneering works of Nusselt\cite{Nusselt1916}, who derived the governing equations for a falling liquid film, and the Kapitza family, who carried out experiments on highly viscous liquids to describe the wave structures in vertically falling liquid films\cite{Kapitza}. The analysis of the instability mechanisms leading to three-dimensional (3D) waves from an initially unperturbed flow has been an active research area since then, involving both theoretical and experimental fluid mechanics\cite{Demekhin2007,Craster2009,Ruyer-Quil2014}.

Until now, most literature has focused on the stability of liquid films falling along inclined and vertical planes. Direct Numerical Simulations (DNS) provide detailed insights into these flows, but the computational cost is unaffordable for most cases. Consequently, simplified numerical models --with different levels of complexity-  have been \textcolor{black}{derived} to explain the vast phenomenology documented in the experiments. These models are simplifications of the Navier-Stoke\textcolor{black}{s} equations. The simplest model is the Benney Equation (BE) \cite{Benney1966}, which assumes that the velocity field \textcolor{black}{is bound to} the thickness evolution. It correctly predicts the instability threshold in inclined and vertical planes but cannot be applied to moderate Reynolds numbers because it blows up at finite times.
On the other hand, Integral Boundary Layer (IBL) models combine the boundary layer approximation with the assumption of self-similarity of the velocity profile to formulate the problem as a function of the flow rate(s), $q_x$ and $q_z$, and thickness, $h$. This approach was first proposed by Kapitza \cite{Kapitza} and Shkadov\cite{Shkadov1970} for stationary and non-stationary waves, respectively, and extended to three-dimensional problems by Shkadov and Demekhin \cite{Demekhin1984}. Unfortunately, although applicable to moderate Reynolds numbers, these models fail to predict the critical Reynolds number in inclined planes. A major modeling improvement in the past decades is the weighted Residuals technique by Ruyer-Quil and Manneville \cite{Ruyer-Quil2000,Ruyer-Quil2002}, which consists in a gradient expansion of the velocity profile to account for high-order non-linearities in the flow.

The flow instabilities that develop on liquid films falling down inclined planes drive an initially unperturbed flat film towards \textcolor{black}{two-dimensional (2D)} solitary waves and 3D wave trains until, finally, reaching a chaotic and unpredictable interface dynamics. Many investigations successfully combined experiments with theoretical models to predict the instability onset and the evolution of the subsequent structures \cite{Alekseenko1985,Liu1994,Ramaswamy1996,Scheid2006}. However, the full spectrum of instability mechanisms is not yet fully understood. The flat film initially undergoes primary instability, leading to 2D structures in the shape of solitary or periodic waves. These interact with each other after a few wavelengths, coalescing or repelling neighboring waves, as shown with noise-driven experiments \cite{Chang1996}. Due to a secondary instability mechanism, these further evolve into three-dimensional structures such as herringbone patterns or solitary 3D waves. A comprehensive experimental investigation is provided by Liu et al. \cite{Liu1995}, who conducted experiments in an inclined plane close to the instability threshold to capture the transition at which the span-wise perturbations are naturally amplified, and the film becomes 3D. On the other hand, Nosoko et al.\cite{Nosoko2004} experimentally introduced artificial span-wise perturbations with an array of needles at the inlet to compare the flow structures with those of unperturbed flows. 

Much less explored is the case of a liquid film dragged by an upward-moving substrate, a configuration found in many coating processes such as dip, curtain, slide, or air-knife coating \cite{Schweizer1997}. In these \textcolor{black}{pre-metered} techniques, the liquid is delivered on a moving substrate, and the resulting flow is known to amplify disturbances \textcolor{black}{of} process parameters, such as the flow rate or the substrate speed\cite{Weinstein2004}. In the case of air-knife coating (also known as jet wiping), in which a high-speed slot gas jet is used to wipe the excess liquid from the substrate, the interaction between the impinging jet and the liquid generates self-sustained perturbations \textcolor{black}{whose frequency depends on the process parameters}\cite{Gosset2019,Mendez2019,Barreiro-Villaverde2021}. 

In all these coating techniques, predicting the downstream evolution \textcolor{black}{of the film} is crucial to ensure the coating uniformity and hence minimize rejection rates due to unmet product quality standards in industrial lines. Coating defects can be damped  --at different rates depending on the scales-- by surface tension and viscosity. This process is known as leveling  \cite{Kheshgi1983,Myers1998}. Orchard \cite{Orchard1963} was the first to estimate the damping of a sinusoidal-shaped coating layer using a linear model that neglects inertial forces. This model was further extended to account for surface tension gradients due to solvent evaporation\cite{Overdiep1986}, to model the effect of surfactants\cite{Schwartz1995}, and to include viscoelastic effects in polymer films\cite{Bousfield1991}. Similarly, Howison et al.\cite{Howison1997} derived a \textcolor{black}{low-dimensional analytical} model that considers variable surface tension, viscosity, solvent diffusivity, and solvent evaporation rate for drying paint layers. Recently, Wang et al.\cite{Wang2019} studied the leveling in bio-based anti-fouling coatings and proposed an empirical model based on the Orchard equation. 

Several studies in the coating community have analyzed the receptivity of the coating film to disturbances to optimize the window of operating conditions. The most common approach uses the Finite Element Method (FEM) because it allows for reproducing the complete geometry and any type of perturbations \cite{Romero2008,Perez2011,Lee2015}. The second option, much less expensive, analyzes the frequency response of an analytical model of the film to small sinusoidal perturbations \cite{Joos1999,Tsuda2010}. In the context of air-knife coating, Mendez et al.\cite{Mendez2021} analyzed the receptivity of the coating layer to several types of gas jet disturbances using an extension of the IBL model for the jet wiping configuration. This analysis was nevertheless limited to a 2D configuration. \textcolor{black}{The stability of falling liquid films subjected to parallel gas flows has been also extensively investigated integral formulations as in Vellingiri et al.\cite{Vellingiri2023}}.

Recently,  Ivanova et al.\cite{Ivanova2022_BLEW3D} applied an IBL model and \textcolor{black}{Linear Stability Analysis (LSA)} to investigate the evolution of 3D perturbations on a vertical and upward-moving flat plate. This configuration turned out to be more stable than falling films because of an additional stabilizing mechanism due to non-linearities acting even when surface tension is neglected.

In this work, we extend the analysis in Ivanova et al.\cite{Ivanova2022_BLEW3D} in various ways. First, we extend the linear stability analysis of the IBL model to a 3D configuration and the non-linear analysis to a broader range of perturbation wave numbers. Second, we validate the IBL model with the pseudo-spectral implementation of surface tension via comparison with the DNS computations. Thirdly, and perhaps more noteworthy, we capture and describe the non-linear and capillary stabilization mechanisms, not only with DNS but also with the reduced order model for the liquid film. %\textcolor{black}{, which has proved to retain most of the physics of the flow at a fraction of the computational cost of the full-scale simulations -> I would remove that}.  %This allowed exploring the limitation of the IBL model and analyzing the origin of the nonlinear stabilization of these flows without the bounds of the usual simplifying hypothesis underpinning the simplified models.

%In this work, we extend the analysis in Ivanova et al.\cite{Ivanova2022_BLEW3D} in various ways. First, we extend the linear stability analysis of the IBL model to a 3D configuration. Second, we extend the nonlinear analysis with a broader range of perturbation wave numbers. Thirdly, and perhaps more noteworthy, we reproduce the same nonlinear stability analysis in the reduced model also to a Direct Numerical Simulation (DNS) of the coating film. This allowed exploring the limitation of the IBL model and analyzing the origin of the nonlinear stabilization of these flows without the bounds of the usual simplifying hypothesis underpinning the simplified models.

The rest of the article is organized as follows: Section~\ref{sec:BLEW} briefly reviews the model formulation, the relevant dimensionless numbers for its scaling, its numerical integration, and the linear stability analysis. Section~\ref{sec:OF} describes the numerical settings for the DNS \textcolor{black}{computations} using the Volume of Fluid (VOF) method. Finally, Section~\ref{sec:results} collects the results, reporting on the stability maps derived from the linear theory, the validation of the integral model for three-dimensional problems with and without surface tension, and the investigation of the role of capillarity and non-linearities on the damping of waves. Conclusions and perspectives are drawn in Section~\ref{sec:conclusions}.

\section{The Integral Boundary Layer Model and its linear stability}
\label{sec:BLEW}

\begin{figure}
    \centering
	\includegraphics[width=0.75\linewidth]{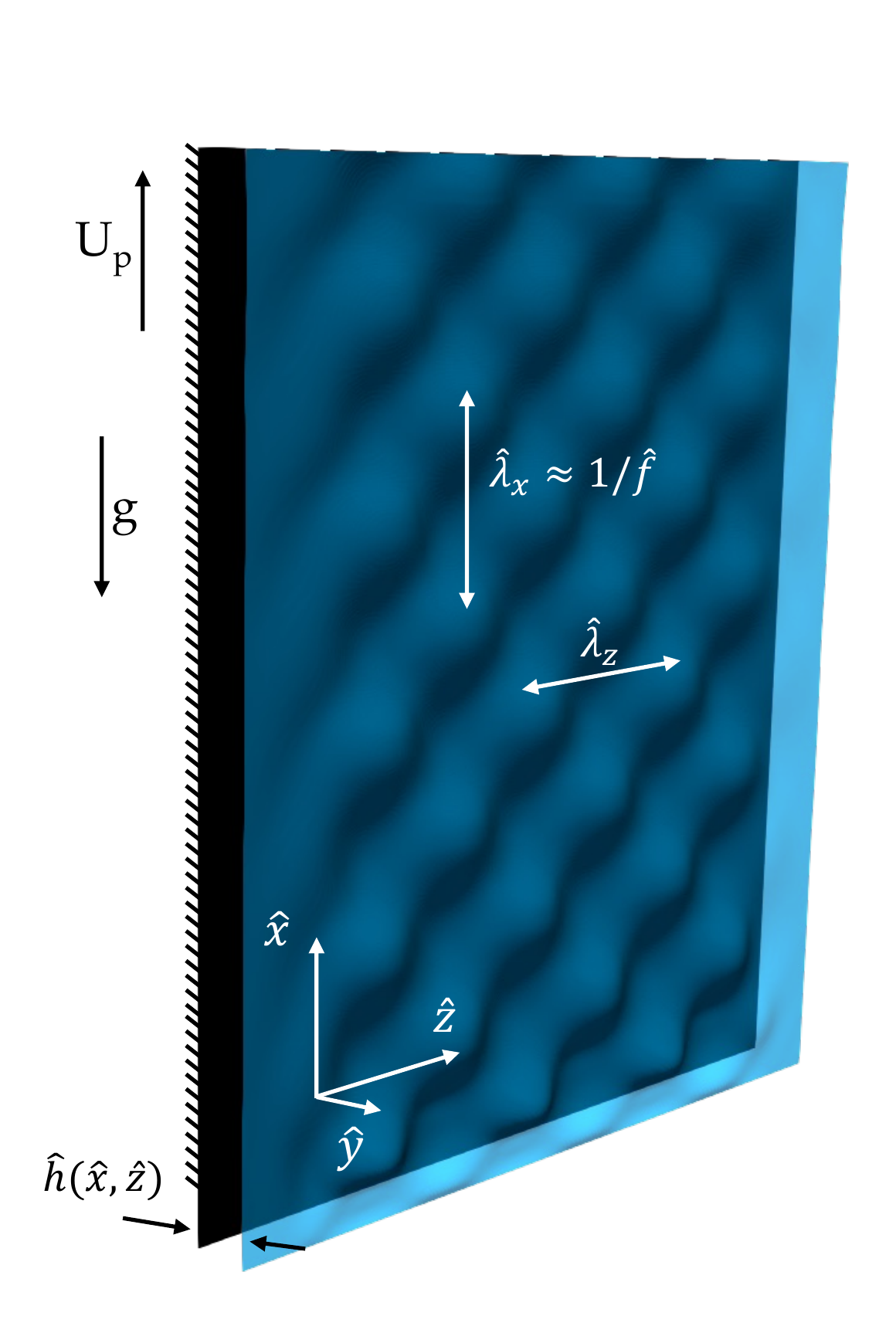} % Here is how to import EPS art
    \caption{ Schematic of the flow configuration investigated in this work with relevant quantities. The reference frame is chosen so that the stream-wise direction $\hat{x}$ is aligned with the substrate (upward) velocity, $U_p$, and opposite to the gravitational acceleration, $g$, although the IBL model is derived considering $x$ in the direction of gravity for consistency with the literature of falling film models.}\label{fig:BLEW_sketch}
 %\caption{ Schematic of the flow configuration investigated in this work with relevant quantities. Note that the reference frame is chosen so that the stream-wise direction $\hat{x}$ is aligned with the gravitational acceleration $g$ and opposite to the substrate (upward) velocity $U_p$.}\label{fig:BLEW_sketch}
\end{figure}

We consider the flow configuration illustrated in Fig.~\ref{fig:BLEW_sketch}: a thin film flow is dragged by a vertical flat surface moving \textcolor{black}{upwards against gravity}. The integral model formulation (IBL) is presented in Section~\ref{sec:BLEW_Equations} while Section~\ref{sec:ST} briefly reports on the numerical methods implemented for its resolution. Section~\ref{sec:BLEW_LSA_equations} presents the linear stability analysis. 

\subsubsection{The IBL model formulation}
\label{sec:BLEW_Equations}

In the vast majority of coating processes, thin films are characterized by considerably larger length scales in the stream wise ($x$) and span-wise ($z$) than in the cross-stream ($y$) directions (cf. Fig. \ref{fig:BLEW_sketch}). In air-knife coating, for example, the film thickness is of the order of 10 microns while the most prominent waves feature a wavelength of the order of a centimeter.

%Therefore, defining as $[h]$ the reference thickness and $[x]$ the reference length to scale the axes as $(\hat{x},\hat{y},\hat{z})=(x/[x],y/[h],z/[x])$, it is possible to introduce a small film parameter $\varepsilon=[h]/[x]<<1$ that weights the contributions of the different forces at play. In this scaling, inertia in the cross-stream coordinate ($y$) is negligible, and pressure can thus be considered constant along $\hat{y}$: this approximation, in nature similar to the one of boundary layer models, is the essence of the long-wave modeling. 

Defining the reference thickness as $[h]$ and the reference length as $[x]$ to scale the axes as $(\hat{x},\hat{y},\hat{z})=(x/[x],y/[h],z/[x])$, it is possible to introduce a film parameter, $\varepsilon=[h]/[x]<<1$, that weights the contributions of the different forces at play. In this scaling, inertia in the cross-stream coordinate ($y$) is negligible, and pressure can thus be considered constant along $\hat{y}$, This approximation is the essence of the long-wave modeling and is similar to the one found in classic boundary layer theory. 

\textcolor{black}{Similar} to the Karman-Polhausen integral formulation in boundary layers, it is possible to integrate the Navier-Stokes equations with the relevant boundary conditions, assuming that the velocity profile remains self-similar within the film. This allows obtaining a \textcolor{black}{2D} reduced order model involving only the film thickness $h(x,z,t)$ and the volumetric flow rates in the stream-wise and span-wise directions, defined as:

\begin{subequations}
\begin{eqnarray}
q_x(x,z,t)=\int^{h(t)}_{0} u(x,y,z,t) dy   \textcolor{black}{\,,}
\end{eqnarray}
and
\begin{eqnarray}
q_z(x,z,t)=\int^{h(t)}_{0} w(x,y,z,t) dy   \textcolor{black}{\,.}
\end{eqnarray}
\end{subequations} 

%Eliminating the velocity field and pressure, a 2D set of equations describes the dynamics of 3D waves. 
The reader is referred to Chang et al. \cite{Chang_book} and Kalliadasis et al. \cite{Kalliadasis2012} for an extensive discussion of these models. In this work, we use the IBL model derived in Ivanova et al. \cite{Ivanova2022_BLEW3D}, nondimensionalized using the scaling discussed in Mendez et al. \cite{Mendez2020}. The set of equations reads:

\begin{subequations}
    \label{eq:BLEW_EQ}
    \begin{eqnarray}
            \partial_{\hat t} \hat h+ \partial_{\hat x }\hat{q}_{\hat{x}} + \textcolor{black}{\partial_{\hat z}  \hat{q}_{\hat{z}}} = 0 \textcolor{black}{\,,}  \label{eq:BLEW_continuity}
    \end{eqnarray} 
    \begin{eqnarray}
         \partial_{\hat t} \hat{q}_{\hat{x}} + \partial_{\hat x} {\mathcal{F}}_{\hat{x}\hat{x}} + \partial_{\hat z} {\mathcal{F}}_{\hat{x}\hat{z}} = \textcolor{black}{\frac{1}{\delta}\Bigl [ \hat h \Bigl ( \partial_{\hat x\hat x\hat x}\hat h  + \partial_{\hat x\hat z\hat z}\hat h +1 \Bigr) + \hat \tau_{w,\hat{x}} \Bigr]} \textcolor{black}{\,,} \label{eq:BLEW_Mx}
    \end{eqnarray}
    \begin{eqnarray}
        \partial_{\hat t} \hat{q}_{\hat{z}} + \partial_{\hat x}{\mathcal{F}}_{\hat{z}\hat{x}}  + \partial_{\hat z} {\mathcal{F}}_{\hat{z}\hat{z}} = \textcolor{black}{\frac{1}{\delta} \Bigl [ \hat h \Bigl ( \partial_{\hat z\hat z\hat z}\hat h + \partial_{\hat z\hat x\hat x}\hat h\Bigr) 
        + \hat \tau_{w,\hat{z}} \Bigr] }  \textcolor{black}{\,,} \label{eq:BLEW_Mz}
    \end{eqnarray}
\end{subequations} where $\delta=\varepsilon Re$ is the reduced Reynolds number, with the film Reynolds number defined as $Re=[h]U_p/\nu_l =({{U}^3_p/g \nu_l})^{1/2}$. The flux terms

\begin{equation}
\mathcal{F}_{\hat{i}\hat{i}} = \int^{\hat{h}(\hat{t})}_0 \hat{v_i} \hat{v_i} dy  
\end{equation} are the integral advective terms, and $\tau_{w,\hat{x}} = -\partial_{\hat{y}} \hat{u} |_{\hat{y}=0}$  and $\tau_{w,\hat{z}}=-\partial_{\hat{y}} \hat{w} |_{\hat{y}=0}$ are the wall shear stresses. The reference quantities used in the scaling of all variables are recalled in Table~\ref{tab:shkadov_scaling}. \textcolor{black}{The long wave assumption leading to the simplified set of equations in \eqref{eq:BLEW_EQ} requires that $\delta \sim \mathcal{O}(1)$, which implies that $Re\sim 1/\varepsilon$.}

\begin{table}
    \caption{Reference quantities, denoted within square brackets, for the Shkadov-like scaling\cite{Mendez2020,Shkadov1970} of the problem. The scaling is designed to keep surface tension at the leading order in the force balance. Hence, the film parameter is $\varepsilon = \mu_l\ U_p/\sigma = Ca^{1/3}$ with $Ca$ the Capillary number.}
    \label{tab:shkadov_scaling}
    \small\addtolength{\tabcolsep}{1.5pt}
    \renewcommand{\arraystretch}{1.5}
    \centering
    \begin{tabular}{c|c|c}
        Reference Quantity & Definition & Expression \\
        $[h]$ & $(\nu_l\,[u]/g)^{1/2}$ & $(\nu_l\,U_p/g)^{1/2}$ \\
        $[x]$ & $[h]/\varepsilon$ & $[h]\,Ca^{-1/3}$ \\
        $[u]$ & $U_p$ & $U_p $ \\
        $[v]$ & $\varepsilon U_p$ & $U_p\,Ca^{1/3} $ \\
        $[p]$ & $\rho_l\,g\,[x]$ & $\rho_l \, g \, [h]\, Ca^{-1/3}$\\
        $[\tau]$ & $\mu_l\,[u]/[h]$ & $(\mu_l\,\rho_l\,g\,U_p)^{1/2}$\\
        $[t]$ & $[x]/[u]$ &  $(\nu_l/U_s\,g)^{1/2}\,Ca^{-1/3}$
    \end{tabular}
\end{table}

Following Demekhin\cite{Demekhin2007}, self-similarity of the velocity profiles is assumed for both components $u(x,y,z)$ and $w(x,y,z)$ to close the model with a relation for the advection and shear stress terms. At first order, these are taken as parabolic profiles obeying the boundary conditions at the wall and the interface. In falling films, the validity of this assumption has been investigated with DNS simulations \cite{Malamataris2002} and experiments \cite{Alekseenko1985}. In air-knife coating, two-phase simulations have shown that the parabolic assumption is reasonable even well outside the asymptotic limit of validity of the long wave assumption\cite{Barreiro-Villaverde2021}. The self-similar velocity profiles are thus expressed as:

\begin{subequations}
    \begin{eqnarray}
        \hat{u} (\hat{h}, \hat{q}_x, \hat{q}_{\hat{z}}) = \frac{3}{2 \hat{h}^3} \Bigl  ( - \hat{h} -\hat{q}_{\hat{x}} \Bigr ) \hat{y}^2 + \frac{3}{\hat{h}^2} \Bigl ( \hat{h} + \hat{q}_{\hat{x}} \Bigr ) \hat{y} - 1  \textcolor{black}{\,,}
    \end{eqnarray} 
    \begin{eqnarray}
        \hat{w} (\hat{h}, \hat{q}_{\hat{x}}, \hat{q}_{\hat{z}}) = \frac{-3 \hat{q}_{\hat{z}} }{2 \hat{h}^3}  \hat{y}^2 + \frac{3 \hat{q}_{\hat{z}}}{\hat{h}^2}  \hat{y}  \,.
    \end{eqnarray}\label{eq:BLEW_velocity}
\end{subequations} 

Introducing \eqref{eq:BLEW_velocity} into the definition of the wall shear stress gives 

\begin{equation}
        \hat{\tau}_{w, \hat{x}}= -\frac{3 \hat{q}_{\hat{x}}}{\hat{h}^{2}} - \frac{3}{\hat{h}}, \,\,\mbox{and}\,\, \hat{\tau}_{w, {\hat{z}}}=-\frac{3 \hat{q}_{\hat{z}}}{\hat{h}^{2}}  \textcolor{black}{\,.}   \label{taus}    
\end{equation} Similarly, introducing \eqref{eq:BLEW_velocity} into the integral advection terms gives

\begin{subequations}
    \begin{eqnarray}
        \mathcal{F}_{{\hat{x}}{\hat{x}}}=  \int_{0}^{\hat{h}} \hat{u}^2 d \hat{y}= \frac{1}{5 \hat{h}} \Bigl( 6 \hat{q}_{{\hat{x}}}^{2} +  2 \hat{h} \hat{q}_{{\hat{x}}} +  \hat{h}^{2} \Bigr) \textcolor{black}{\,,}
    \end{eqnarray}

    \begin{eqnarray}
        \mathcal{F}_{{\hat{x}}{\hat{z}}}=\mathcal{F}_{{\hat{z}}{\hat{x}}}= \int_{0}^{\hat{h}} \hat{u} \hat{w} d \hat{y}= \frac{1}{5 \hat{h}} \Bigl( 6 \hat{q}_{{\hat{x}}} \hat{q}_{{\hat{z}}}  +  \hat{h} \hat{q}_{{\hat{z}}}  \Bigr) \textcolor{black}{\,,}
    \end{eqnarray}

    \begin{eqnarray}
        \mathcal{F}_{{\hat{z}}{\hat{z}}}= \int_{0}^{\hat{h}} \hat{w}^2 d \hat{y}= \frac{6 \hat{q}_{{\hat{z}}}^{2}}{5 \hat{h}} \textcolor{black}{\,.}
    \end{eqnarray}  \label{eq:BLEW_F} 
\end{subequations} 

The system of equations \eqref{eq:BLEW_continuity}-\eqref{eq:BLEW_Mz} is closed by relations \eqref{taus} and \eqref{eq:BLEW_F}.

To give an order of magnitude of the dimensionless quantities encountered in standard coating processes, Table~\ref{tab:references_coating} collects the typical values of dimensional and dimensionless parameters of interest and the wave number $k=2 \pi/\lambda$ of the most commonly observed perturbations. This study will focus on those ranges to provide conclusions directly relevant to coating processes. \textcolor{black}{It is worth noticing that the range of reduced $\delta$ is well above the limit $\delta \sim \mathcal{O}(1)$ that justifies the asymptotic expansion supporting the consistency of the model in Eq.\ref{eq:BLEW_EQ}. Therefore, the fact that the IBL model prediction is in agreement with the DNS, as later showcased, came as an unexpected and noteworthy surprise.}

\begin{table*}
\caption{Reference quantities of different coating processes according to the bibliography.}
\label{tab:references_coating}
\centering
\begin{tblr}{
  column{even} = {c},
  column{3} = {c},
  column{5} = {c},
  column{7} = {c},
  column{9} = {c},
  vline{2} = {-}{},
  hline{1-2,6} = {-}{},
}
Coating method           & ${h}_0 [\mu m]$ & $\hat{h}_0$ & $U_p$ [m/s] & $\delta$ & $Ca$        & $\lambda [mm]$ & $\hat{k}_x$ & $f$ [Hz]   & $\hat{f}$ \\
Air-knife coating \cite{Gosset2019,Mendez2019,Mendez2021}       & 5.0-25        & 0.02-0.12    & 0.5-2       & 70-300   & 0.003-0.01 & 3-30         & 0.3-2       & 40-500  & 0.05-0.3   \\
Slot die coating \cite{Romero2008,Perez2011,Lee2015,Tsuda2010,Silva2023}   & 10-100        & 0.02-0.3    & 0.05-2      & 1-260    & 0.05-4      & 1.5-3.5        & 1.5-4.5     & 30-80    & 0.2-0.7   \\
Slide die coating  \cite{Jiang2005}    & 70-150        & 0.02-0.12   & 0.1-1.5     & 1-120    & 0.1-1.5     & 0.3-0.8        & 10.0-32     & 100-2000 & 1.5-5     \\
Gravure roll coating \cite{Kapur2003} & 3.0-7         & 0.01-0.03   & 0.16-2      & 3-300    & 0.01-0.1    & 0.15-0.5       & 7.0-40      & 400-2600 & 1.0-7     
\end{tblr}
\end{table*}

\subsubsection{Numerical Methods for the IBL integration}
\label{sec:ST}

The IBL model was integrated in this work using the Finite Volume (FV) BLEW (Boundary LayEr Wiping) code developed at the von Karman Institute \cite{Mendez2018b,Mendez2021,Ivanova2022_BLEW3D}. The reader is referred to Ivanova et al\cite{Ivanova2022_BLEW3D} for a description of the FV formulation and related time-stepping. \textcolor{black}{The latter} combines the two-step second-order Lax-Wendroff and the two-step first-order Lax-Friedrich schemes using a user-specified blending (e.g.: 30 \% Lax-Wendroff and 70 \% Lax-Friedrich) or flux/slope limiters such as minmod, van Albada, and others\cite{LeVeque2002}. %The impact of the numerical setup of the simulation is further discussed in Section~\ref{sec:validation}.

In all the investigated test cases, we consider a rectangular computational domain, introducing the disturbances at the bottom boundary (cf. Fig.~\ref{fig:BLEW_sketch}). The perturbations are introduced in the stream-wise flow rate as 

\begin{equation}
\label{eq:3Dperturbation}
    \hat{q}_{x} (0,\textcolor{black}{\hat{z},\hat{t}}) = \hat{q}_{x0} \Biggl[1 + A \sin\bigl(2 \pi \hat{f} \hat{t}\bigr )\sin \Bigl(\frac{2 \pi}{\hat{\lambda}_z} \Bigr) \Biggr] l(\textcolor{black}{\hat{z}}) \textcolor{black}{\,,}
\end{equation} where $A$ is the relative amplitude of the perturbation with respect to the undisturbed base state, $\hat{f}$ is the dimensionless frequency (related to the stream-wise dimensionless wavelength $\hat{\lambda}_x$ through the advection velocity $\approx 1$), $\hat{\lambda}_z$ is the dimensionless span-wise wavelength, and $l(\hat{z})$ is the smooth step function acting as a mask to \textcolor{black}{damp} the disturbances to zero on the lateral sides of the domain, hence limiting the waves to the range $\hat{\lambda}_z / 2 < \hat{z} < \hat{L}_z - \hat{\lambda}_z / 2$. The initial (unperturbed) conditions are given by the steady state solution of the IBL model, i.e. :

\begin{equation}
\label{eq:base_state}
\begin{cases}
\hat{h}(\hat{x},\hat{z},0) = \hat{h}_0 \textcolor{black}{\,,} \\
\hat{q}_{x}(\hat{x},\hat{z},0) = \frac{1}{3}\hat{h}_{0}^3-\hat{h}_{0} \textcolor{black}{\,,}\\
\hat{q}_{z}(\hat{x},\hat{z},0) = 0	\textcolor{black}{\,.}	\\
\end{cases}
\end{equation}

The numerical domain covers a distance $\hat{L}_x=5\lambda_x$ in the stream-wise direction $\hat{x}$, and $\hat{L}_z=4\lambda_z$ in the span-wise direction $\hat{z}$. The computational mesh is uniform and square ($\Delta x = \Delta_z=\Delta$), with a cell size adjusted to the perturbation frequency using a user-defined parameter $n_{\lambda}$, such that the number of cells per dominant wavelength is $\Delta = min(1/\hat{f}, \hat{\lambda}_z) / n_{\lambda}$. All boundaries, except the bottom one, are treated as outlets using linear extrapolation. 

An important extension over the previous implementation of the solver lies in the treatment of surface tension. In literature, this term is usually implemented using Fourier Spectral methods (FSM)\cite{Tseluiko2011,Scheid2006,Dietze2013,Denner2018}, which provide high accuracy and negligible numerical diffusion. However, this implementation requires periodic boundary conditions, e.g. $h(x=0) = h(x=L_x)$. This type of configuration is known as \textit{closed flow} condition, and its application is not possible in \textit{open flow} settings where perturbations are continuously introduced on one side of the domain and left growing unbounded downstream \cite{Ramaswamy1996}.

With the aim of balancing the accuracy of spectral-based computation of the surface tension term with the robustness of the available FV formulation, we implemented in this work a filtered spectral differentiation for the third derivatives in Eq.~\ref{eq:BLEW_Mx} and \ref{eq:BLEW_Mz}. Let $H(\textcolor{black}{\hat{k}_x,\hat{k}_z, \hat{t}}$) denote the 2D Discrete Fourier Transform of the film thickness $\hat{h}(\hat{x},\hat{z},\textcolor{black}{\hat{t}})$ at time $\textcolor{black}{\hat{t}}$, that is:

\begin{equation}
	H(\hat{k}_x, \hat{k}_z, \hat{t})= \mathcal{F}\{\hat{h}\}= \sum_{n=0}^{n_x-1} \sum_{m=0}^{n_z-1} \hat{h}(\hat{x}_n, \hat{z}_m) e^{ -\textcolor{black}{\mathrm{i}} (\hat{k}_{x} \hat{x}_n + \hat{k}_{z} \hat{z}_m)}  \textcolor{black}{\,,}
\end{equation} where $\hat{k}_x$ and $\hat{k}_z$ are the wave numbers in each direction, $\textcolor{black}{\mathrm{i}}=\sqrt{-1}$ , and $n_x$ and $n_z$ are the number of grid points in $\hat{x}$ and $\hat{z}$, respectively. We denote as $\mathcal{F}^{-1}()$ the inverse discrete Fourier transform, such that $\hat{h}=\mathcal{F}^{-1}(H)$.

Let $H_F(\hat{k}_x,\hat{k}_z)$ denote the 2D transfer function of a second-order isotropic Butterworth filter:

\begin{equation}
    H_F(\hat{k}_x , \hat{k}_y) = 1 / [1 + (D(\hat{k}_x, \hat{k}_y) / D_0)^4]  \textcolor{black}{\,,}
\end{equation} where $D=(\hat{k}_x^2 + \hat{k}_z^2)^{1/2}$ and $D_0= min(1/\hat{f}, \hat{\lambda}_z) F_{cut}$ is the cutoff wave number, based on the shortest wavelength introduced by the flow rate perturbation, and $F_{cut}$ a user-defined cut-off parameter that modulates the filtering intensity.  Its effect on the computations is addressed later in Section~\ref{sec:validation}.

The low-pass filtered signal of the film thickness is thus the inverse of the product $\tilde{h}=\mathcal{F}^{-1}(H H_{\textcolor{black}{F}})$. The smoothed derivatives can be easily computed in the Fourier domain. For example, the smoothed capillary terms in Eq.~\ref{eq:BLEW_Mx} becomes 

\begin{equation}
\partial_{\hat{x}\hat{x}\hat{x}} \hat{h}= \mathcal{F}^{-1}(\hat{k}^3_x H H_{\textcolor{black}{F}})\,\,\mbox{and}\,\, 
\partial_{\hat{z}\hat{x}\hat{x}} \hat{h}= \mathcal{F}^{-1}(\hat{k}_z\hat{k}^2_x H H_{\textcolor{black}{F}}) \textcolor{black}{\,.}
\end{equation}

\begin{figure}
    \centering
	\includegraphics[width=\linewidth]{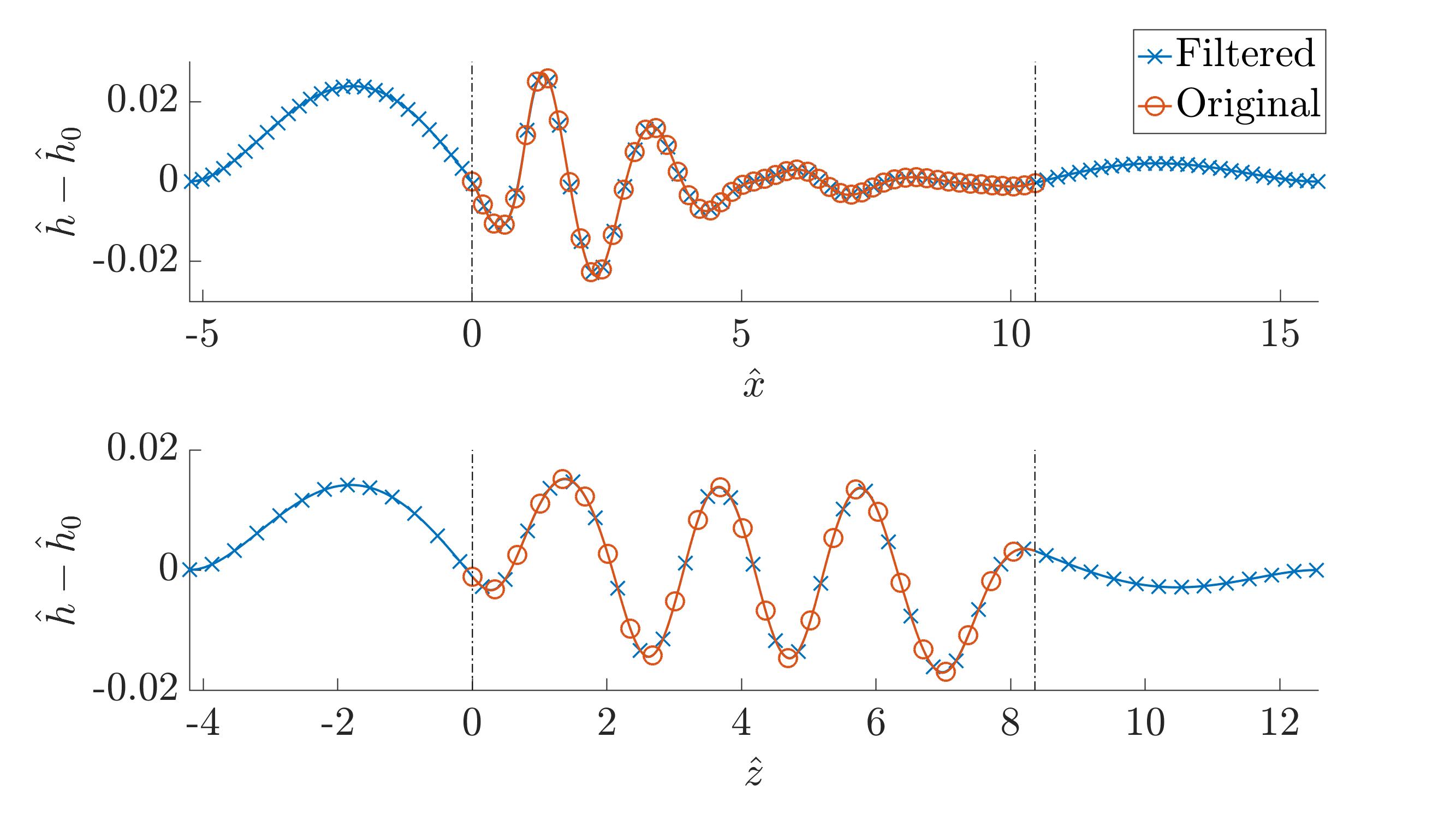} % Here is how to import EPS art
	\caption{\textcolor{black}{Original thickness profiles along the stream-wise $\hat{x}$ (top) and span-wise $\hat{z}$ (bottom) coordinates together with the smoothed version $\mathcal{F}^{-1}(H H_{\textcolor{black}{F}})$ used to compute the capillary terms. Outside the black dashed lines, the original thickness profiles are extrapolated and periodicity is enforced at both sides. The data correspond to the validation test case with surface tension in Section~\ref{sec:validation}}.}\label{fig:extension_spectral}
 % DAVID: do not spend too much time on this; a figure of the h profile is largely sufficient.
\end{figure}

Since the aforementioned spectral computation implies the periodicity of $\hat{h}$, we artificially extend the domain as shown in Fig.~\ref{fig:extension_spectral} using a cubic extrapolation, and then truncate it back to its original length. This is equivalent to a linear extrapolation of the third derivatives in the surface tension term and mitigates problems at the boundaries due to the lack of periodicity. \textcolor{black}{A domain extension of $L_x/4$ and $L_z/4$ in $x$ and $z$ respectively proved sufficient in all cases, although this parameter depends largely on the domain discretization, numerical methods and wavelengths among others.}

The spatial filtering in the computation of the capillary terms is here considered for numerical stabilization purposes, \textcolor{black}{in particular} to limit numerical dispersion. The rationale is to avoid the third derivatives interacting with wavelengths that would be physically damped but might arise because of numerical artifacts.

\subsubsection{Linear stability analysis (LSA) of the IBL model}
\label{sec:BLEW_LSA_equations}

We now extend the linear stability analysis in Ivanova et al.\cite{Ivanova2022_BLEW3D} to three-dimensional perturbations. We are thus interested in the linearization of \eqref{eq:BLEW_EQ}-\eqref{eq:BLEW_F} around the steady-state solution and introduce a perturbation on the flow variables $(h,q_x,q_z)$. Since all variables are dimensionless in the following, we drop the $\hat{}$ notation for simplicity. Given $\phi=\phi_0+\tilde{\phi}$ the decomposition of the general variable $\phi$ into the base state $\phi_0$ and the perturbation $\tilde \phi\ll 1$, we consider a normal \textcolor{black}{mode} perturbation $\tilde{\phi} = \phi_{\epsilon} e^{\textcolor{black}{\mathrm{i}}(\mathbf{k}\cdot\mathbf{x} - \omega t)}$ with $\mathbf{k}=[k_x,k_z]\in\mathbb{R}^2$ the wave number vector, $\mathbf{x}=[x,z]$, $\omega=\omega_r+\textcolor{black}{\mathrm{i}}\omega_i\in \mathbb{C}$ the angular frequency. Introducing the harmonic perturbation, linearizing the set of equations, and simplifying the terms related to the base state leads to:

	\begin{subequations}
		\label{eq:stability_BLEW}
		\begin{eqnarray}
				-i \omega h_{\epsilon} + i k_x q_{x,\epsilon} + i k_z q_{z,\epsilon} = 0 \textcolor{black}{\,,}\label{eq:stability_mass}
		\end{eqnarray}
		\begin{multline}
		     - i h_0^2 \omega q_{x,\epsilon} + i k_x \Biggl[ \frac{2 q_{x,\epsilon} }{5} \bigl( q_{x0} h_0 + h_0^2 \big) + \frac{h_{\epsilon}}{5} \big(  h_0^2 - 6 q_{x0}^2 \big) \Biggr] \\ \
       + i k_z h_0 \frac{q_{z,\epsilon}}{5}  \big( 6 q_{x0} + h_0 \big)  \\
      = \frac{1}{\delta} \Bigl[ 3 h_0 ^2 h_{\epsilon} -  i h_0 ^3 k_x h_{\epsilon} \big( k_x ^2 + k_z ^2 \big) - 3 q_{x,\epsilon} - 3 h_{\epsilon} \Bigr]   \textcolor{black}{\,,} \label{eq:stability_momX}
		\end{multline}
		\begin{multline}
		     - i h_0^2 \omega q_{z,\epsilon} + i k_x \Biggl[ \frac{q_{z,\epsilon}}{5} \big( 6 q_{x0} h_0 + h_0^2 \big) \Biggr] \\
       = \frac{1}{\delta} \Bigl[ - i h_0 ^3 k_z h_{\epsilon}  \big( k_x ^2 + k_z ^2 \big) - 3 q_{z,\epsilon} \Bigr]\,.  \label{eq:stability_momZ}
		\end{multline}
	\end{subequations}

This set of equations constitutes the dispersion relation $\mathcal{D}$ of the frequencies $\omega\in\mathbb{C}$ to the wave numbers $\mathbf{k}$ for a given set of parameters $(h_0,\delta)$, i.e. $\mathcal{D}(\mathbf{k},\omega; h_0,\delta)=0$. For a given pair $(h_0,\delta)$, the system is \emph{linearly stable} against perturbations with wave numbers $\mathbf{k}$ if the associated $\omega\in\mathbb{R}$ has $\textcolor{black}{\omega_i}=\mbox{Im}(\omega)<0$ and \emph{linearly unstable} if $\textcolor{black}{\omega_i}=\mbox{Im}(\omega)>0$. Solutions with $\textcolor{black}{\omega_i}=0$ are \emph{neutrally stable}.

The mapping of the stability regions for each pair $(h_0,\delta)$ was carried out by solving the equations \eqref{eq:stability_BLEW} for $\omega_r,\textcolor{black}{\omega_i}$ over a wide range of wave numbers $\mathbf{k}$. The equations were solved using the \textit{fsolve} function in the SciPy\cite{Scipy} package.

\section{Two-phase flow CFD simulations}
\label{sec:OF}

The Direct Numerical Simulations (DNS) of the liquid film flow were carried out with interFoam solver in OpenFOAM. Section~\ref{sec:OF_VOF} presents the governing equations based on the two-phase flow problem formulation; Section~\ref{sec:OF_setup} presents the simulation setup.

% DAVID: Do not play self-defensive. You do not do things because you can't do others (due to the lack of experiments, we do this.) You do what you do and you write about it ;)

\subsection{Volume of Fluid formulation}
\label{sec:OF_VOF}

% DAVID: I use the same notation as before for the flow velocities. No need to use the capital letter here since this is DNS (no LES and no RANS)

The simulations were carried out using an algebraic formulation of the Volume Of Fluid (VOF) \cite{Hirt1981} method in which the interface capturing is based on an extra variable, known as the liquid volume fraction $\alpha$. The value at a cell is $\alpha=1$ if it is filled with phase 1 (liquid), and $\alpha=0$ if it is filled with phase 2 (gas). The interface lies in the cells where $0 < \alpha < 1$. 

The volume fraction is transported by 

\begin{equation}
\label{eq:advection_alpha_org}
    \frac{\partial \alpha}{\partial t} +  \nabla \cdot (\alpha \,\mathbf{u})  + \nabla \cdot ( \mathbf{u}_r \alpha (1 - \alpha )) = 0 \textcolor{black}{\,,}
\end{equation} where $\mathbf{u}=(u,v,w)$ is the flow velocity, and the last term includes an artificial velocity $\mathbf{u}_r$ to limit additional diffusion produced by the interface smearing. This term is based on the face flux, and its magnitude can be adjusted with the parameter $c_\alpha$, which ranges between 0 (compression deactivated) and 1 (maximum compression). \textcolor{black}{The} equation is solved using the Multidimensional Universal Limiter for Explicit Solution (MULES) algorithm, an OpenFOAM implementation of the flux-corrected transport technique that guarantees boundedness in hyperbolic equations. The reader interested in a complete description of the implementation and the validation is referred to Márquez Damián\cite{MarquezDamian2013}, and Deshpande et al.\cite{Deshpande2012}.

The system of PDEs is completed with the incompressible Navier-Stokes equations:

\begin{subequations}
\begin{eqnarray}
\label{eq:N_S_1}
\nabla \cdot \mathbf{u} = 0 \textcolor{black}{\,,}
\end{eqnarray}
\begin{eqnarray}
\label{eq:N_S_2}
  \frac{\partial \left( \rho \mathbf{u} \right)}{\partial t} + \nabla \cdot \left( \rho \mathbf{u} \otimes \mathbf{u} \right)   =  - \nabla p + \rho \mathbf{g}  + \nabla \cdot [ 2\mu \mathbf{S}]  + S_\sigma \textcolor{black}{\,,}
%\rho \frac{D \mathbf{U}}{Dt} =  - \nabla p + \rho g + \nabla \cdot [ 2(\mu + \mu_t) \mathbf{S}]  + S_\sigma 
\end{eqnarray}
\end{subequations} where $p$ is the pressure, $\mu$ is the dynamic viscosity, $\mathbf{S}=(\nabla \mathbf{u} + \nabla \mathbf{u}^T)/2$ is the strain rate tensor and $S_\sigma$ is the term accounting for the capillary pressure induced by surface tension. The latter is computed using the Continuum Surface Force (CSF) model. The fluid properties $\phi$ are locally computed using a weighted average ($\phi = \phi_1 \alpha + \phi_2 (1-\alpha)$) based on the volume fraction $\alpha$ and the fluid properties of the phases ($\phi_{1}$ and $\phi_{2}$).

\subsection{Numerical setup}
\label{sec:OF_setup}

\begin{figure}
    \begin{subfigure}{0.45\textwidth}
      \includegraphics[width=\linewidth]{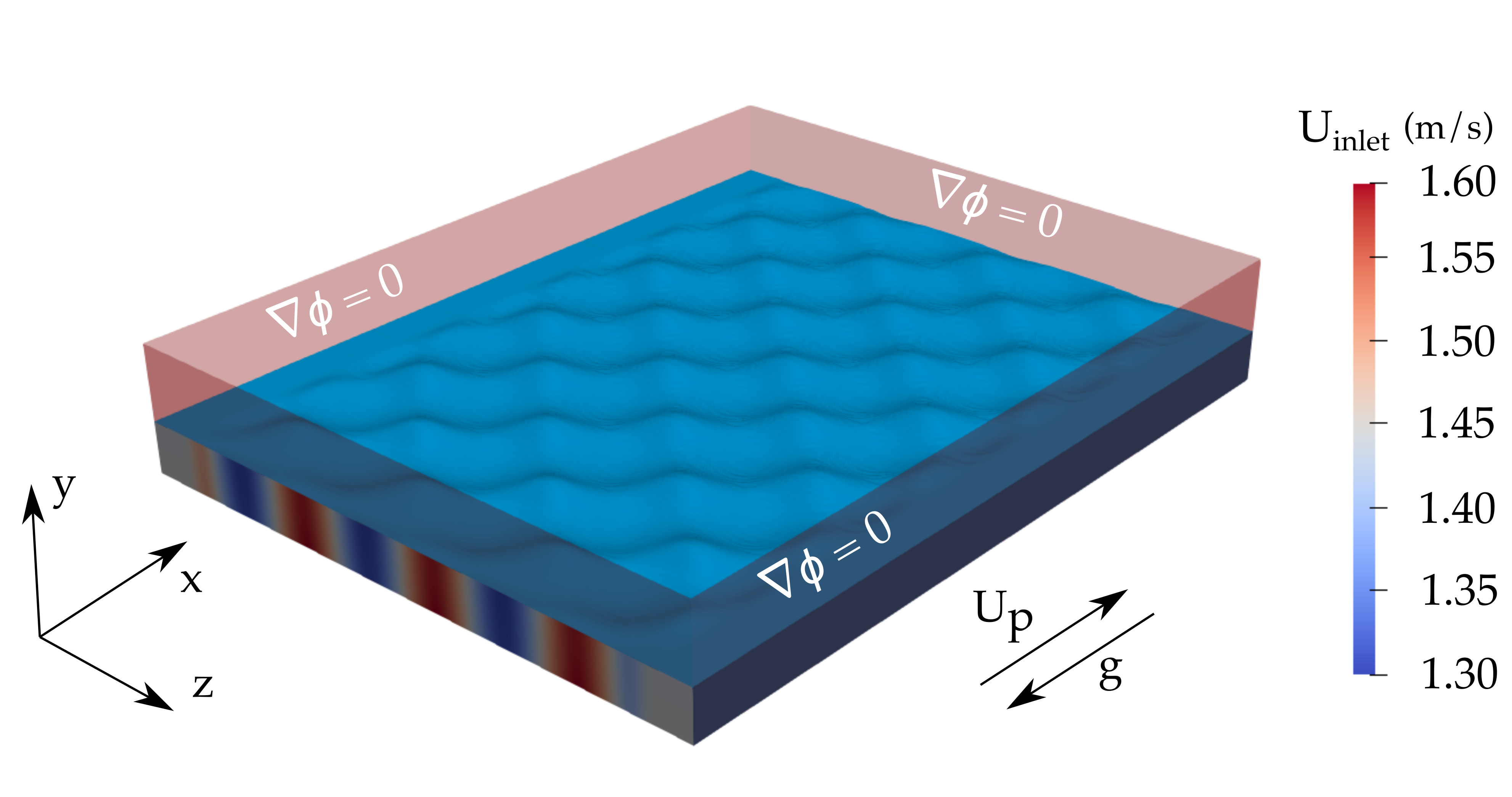}
      \caption{Three-dimensional representation of the computational domain of the two-phase DNS simulations.}
      \label{fig:3D_domain_OpenFOAM}
    \end{subfigure}

    \medskip
    
    \begin{subfigure}{0.45\textwidth}
      \includegraphics[width=\linewidth]{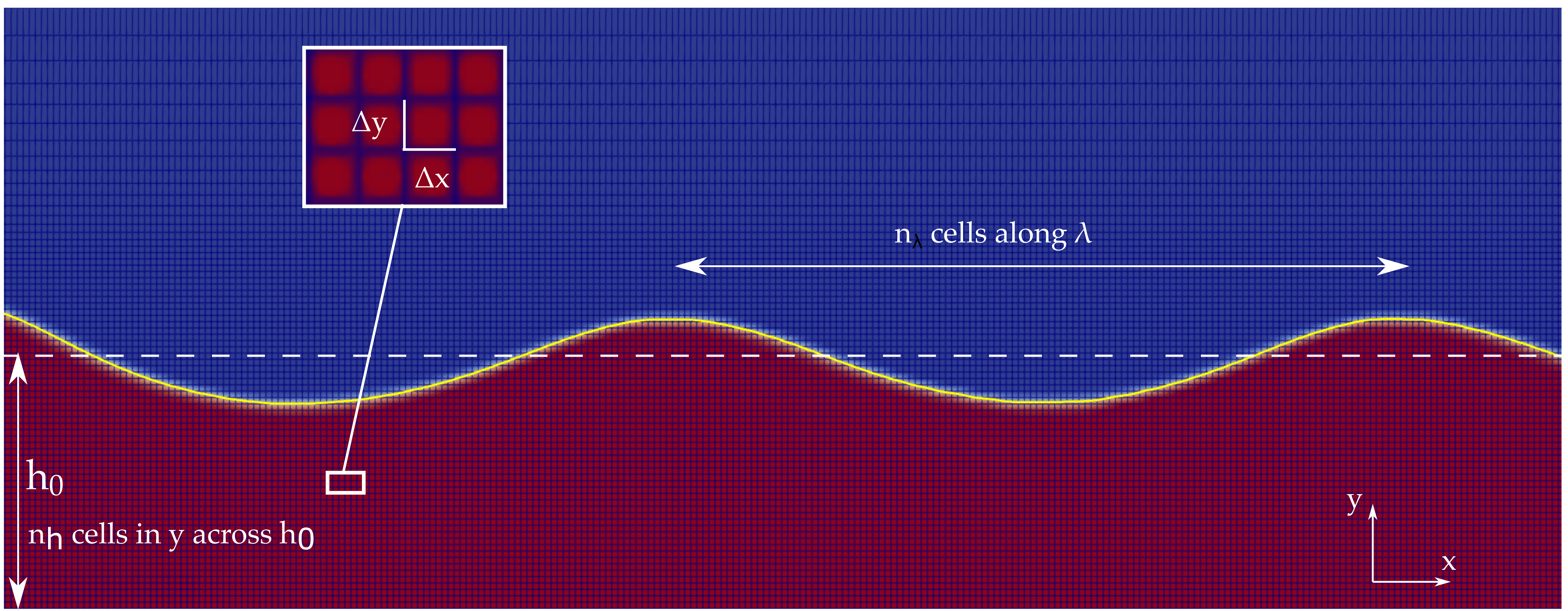}
      \caption{Slice of the mesh taken at a z-midpoint.}
      \label{fig:2D_mesh_OpenFOAM}
    \end{subfigure}
\caption{\label{fig:OpenFOAM_domain} Computational domain (a) and mesh (b) for the DNS simulations. The boundary conditions represented in (a) are: a wall on the bottom, where the velocity, $U_p$, is opposite to gravity, $g$. The red boundaries are outlets where a zero gradient condition is set for all variables, and the boundary on the top (not shown in this figure) fixes the reference pressure $p=0$. The inlet is colored with the velocity field perturbed as described in Eq.~\ref{eq:3Dperturbation} and with the velocity profile across the film given by Eq.~\ref{eq:BLEW_velocity}. The y-coordinate is scaled by a factor 25 for plotting purposes.} 
\end{figure}

 The computational domain for the DNS simulations is represented in Figure~\ref{fig:OpenFOAM_domain} together with the relevant boundary conditions. Its extension in the stream-wise and cross-stream-wise directions is the same as the one in the IBL simulations with the code BLEW. The inlet is split into two patches: one for the liquid film (below) and one for the ambient air (above). The flow is perturbed in terms of the liquid velocity field \textcolor{black}{instead of on the flow rate as in the IBL solver}. A velocity profile must therefore be prescribed at the inlet. Both parabolic and constant profiles were considered, but the impact on the wave dynamics was found to be negligible.  Finally, the patch corresponding to the substrate is defined as a wall with velocity $U_p$ while the other patches are configured as outlets (zero gradients).

The mesh size was adjusted using the same approach than in BLEW, based on the number of cells per \emph{expected} wavelength. In the cross-stream direction, the grid spacing is $\Delta_y = h_0 / n_h$, where $n_h$ is the number of cells across the film thickness $h_0$, until $y=2 h_0$. From here, a  cell-to-cell expansion ratio of 1.1 is applied until the top boundary, located at $y=2.5 h_0$ where the reference pressure is fixed ($p=0$).  
 
The simulations were conducted with water ($\rho_l = 1000$ kg/m$^3$, $\mu_l= 0.001$ Pa$\cdot$s and $\sigma = 0.0728$ N/m) as a working liquid, and air ($\rho_g = 1.2$ kg/m$^3$ and $\mu_g = 1.776 \times 10^{-5}$ Pa$\cdot$s). These were initialized with a liquid film of constant thickness $h = \hat{h}_0 [h]$ and \textcolor{black}{uniform} velocity equal to the substrate $U_p$. The Crank-Nicolson temporal scheme with a blending parameter 0.5 was used for the time stepping; this provided a reasonable compromise between accuracy and robustness. 
The sensitivity of the results to the mesh discretization is examined in Section~\ref{sec:validation}.

\section{Results}
\label{sec:results}

The results are divided into three sections. Section~\ref{sec:lin_stability} presents the results of the LSA of the IBL model. Section~\ref{sec:validation} reports on the sensitivity of the IBL model to finite amplitude perturbations and its validation with the DNS computations. Finally, the section closes in Section~\ref{sec:results_nonlinear} with a discussion on the role of capillarity and non-linearities on the damping of three-dimensional waves. 

\subsection{Linear stability of the IBL model}
\label{sec:lin_stability}

\textcolor{black}{We consider the solution $\omega_i$ of the dispersion relation in Section~\ref{sec:BLEW_LSA_equations} for several sets of parameters $[\delta,\hat{h}_0]$ over a range of dimensionless wave numbers $\mathbf{\hat{k}}=(\hat{k}_x,\hat{k}_z)$}. \textcolor{black}{The latter are compatible with the process windows of several coating techniques analysed in the literature\cite{Gosset2019,Mendez2019,Mendez2021,Romero2008,Perez2011,Lee2015,Tsuda2010,Jiang2005,Kapur2003}.}. \textcolor{black}{We mainly focus on air-knife and slot die coating; hence, we restrict the analysis to three reduced Reynolds numbers ($\delta$ = 50, 150, and 250) and three dimensionless thicknesses ($\hat{h}_0$ = $0.05$, $0.1$, $0.2$). The selected range of dimensionless wave numbers $k_x,k_z\in [0,5]$ covers a wide range of operating conditions according to Table~\ref{tab:references_coating}. The highest wave number corresponds to a dimensionless frequency of $\hat{f}\approx 0.8$, which is a significantly high considering than the frequency of maximum receptivity in 2D air-knife coating was found to be at approximately $\hat{f}=0.05$\cite{Mendez2021}. In slot-die coating, however, the receptivity of the film is larger and the wave numbers rise up to $\hat{k}_x$ of 4.5.}

\begin{figure*}[]
    \centering % <-- added

\begin{subfigure}{0.33\textwidth}
  \includegraphics[width=\linewidth]{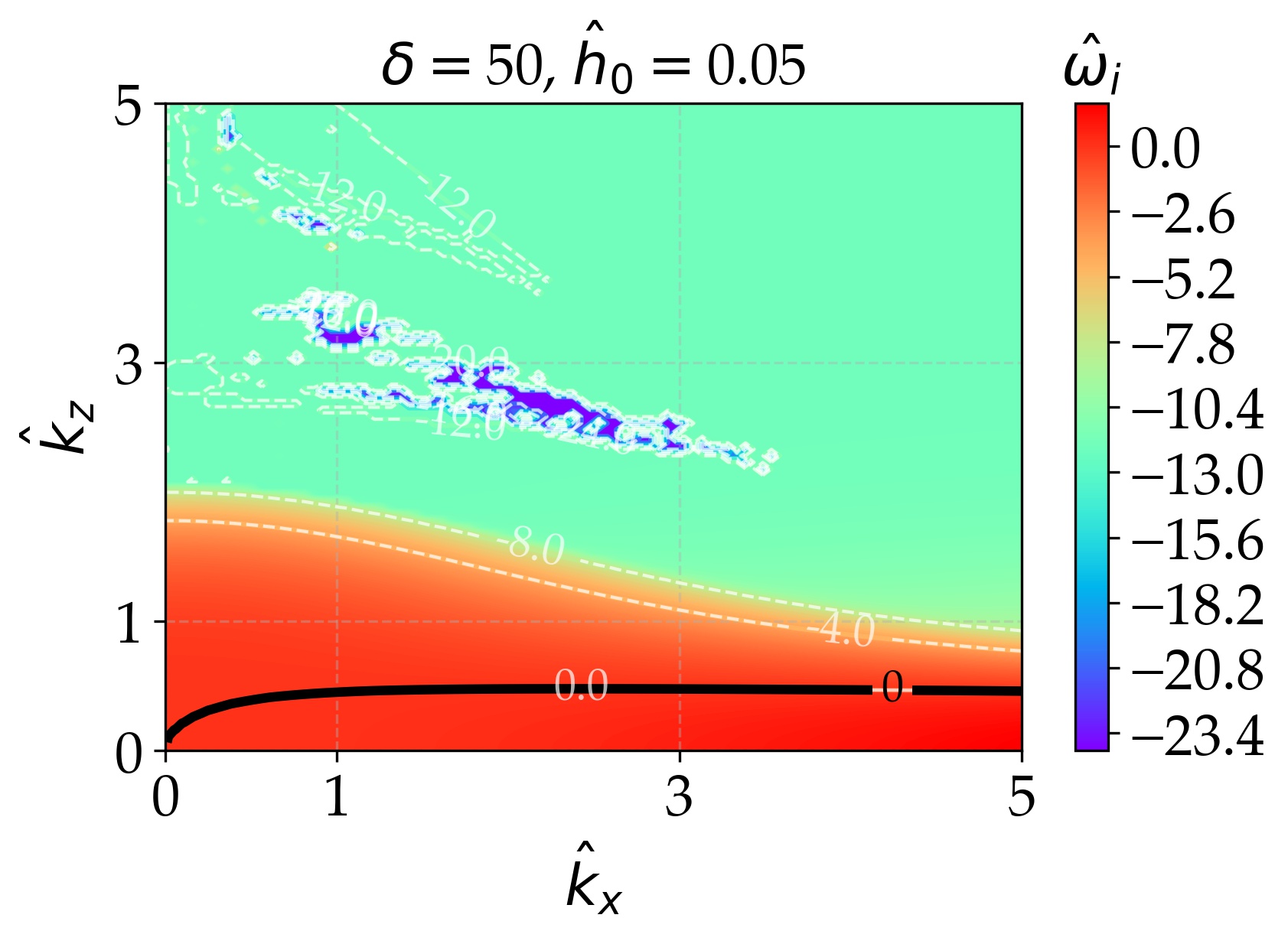}
  \caption{}
  \label{fig:h005_delta50}
\end{subfigure}\hfil % <-- added
\begin{subfigure}{0.33\textwidth}
  \includegraphics[width=\linewidth]{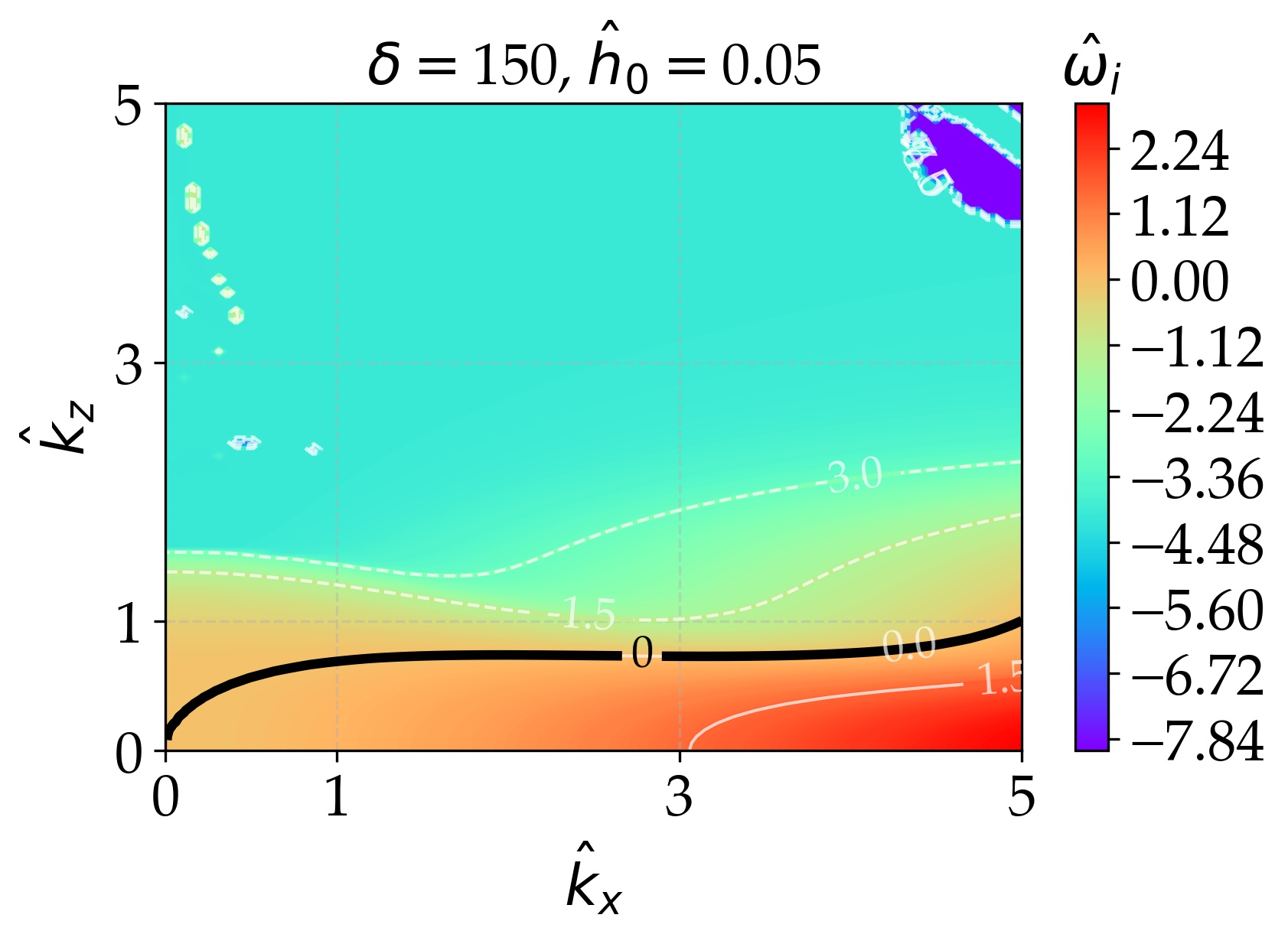}
  \caption{}
  \label{fig:h005_delta150}
\end{subfigure}\hfil % <-- added
\begin{subfigure}{0.33\textwidth}
  \includegraphics[width=\linewidth]{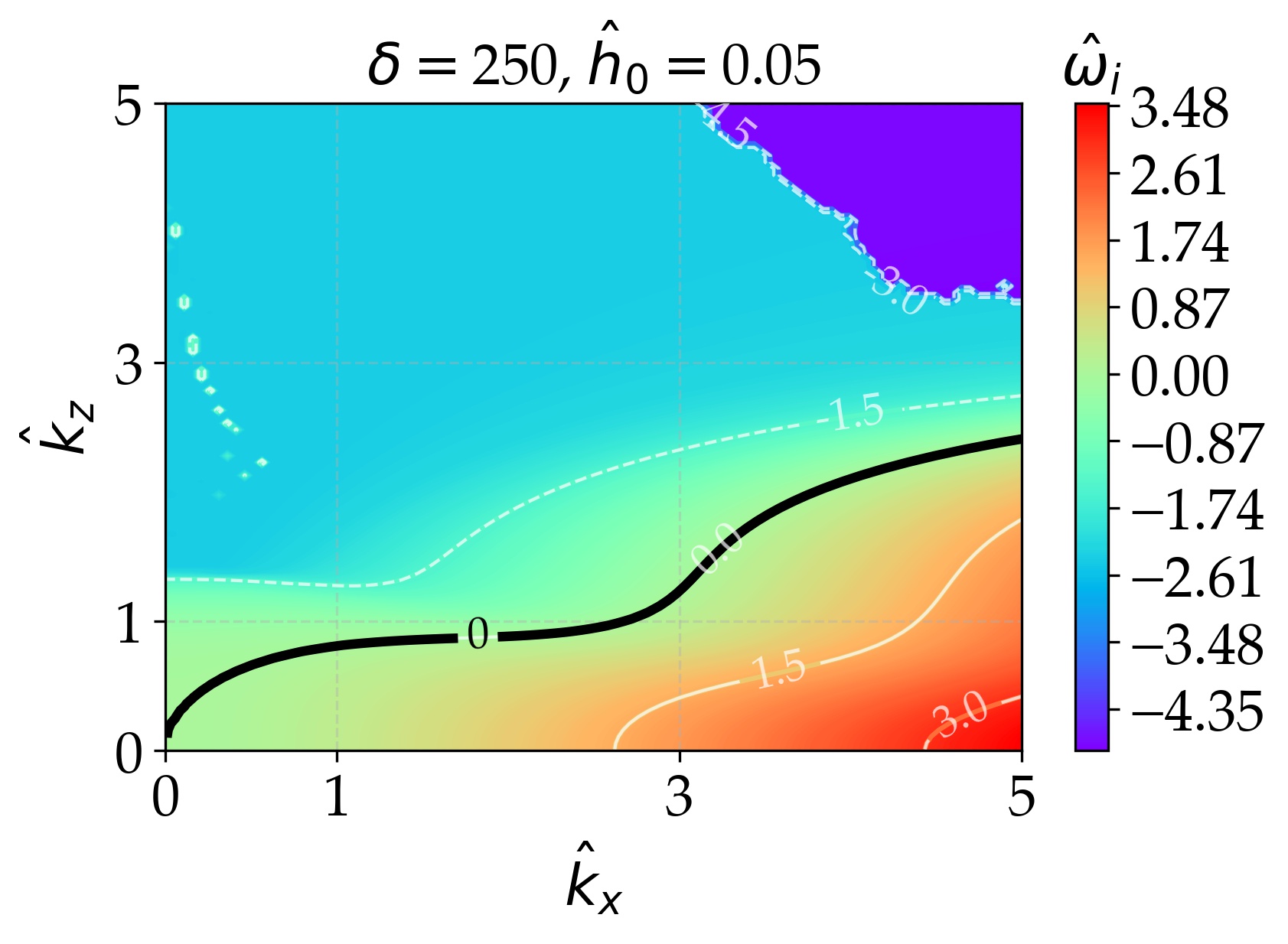}
  \caption{}
  \label{fig:h005_delta250}
\end{subfigure}

\medskip

\begin{subfigure}{0.33\textwidth}
  \includegraphics[width=\linewidth]{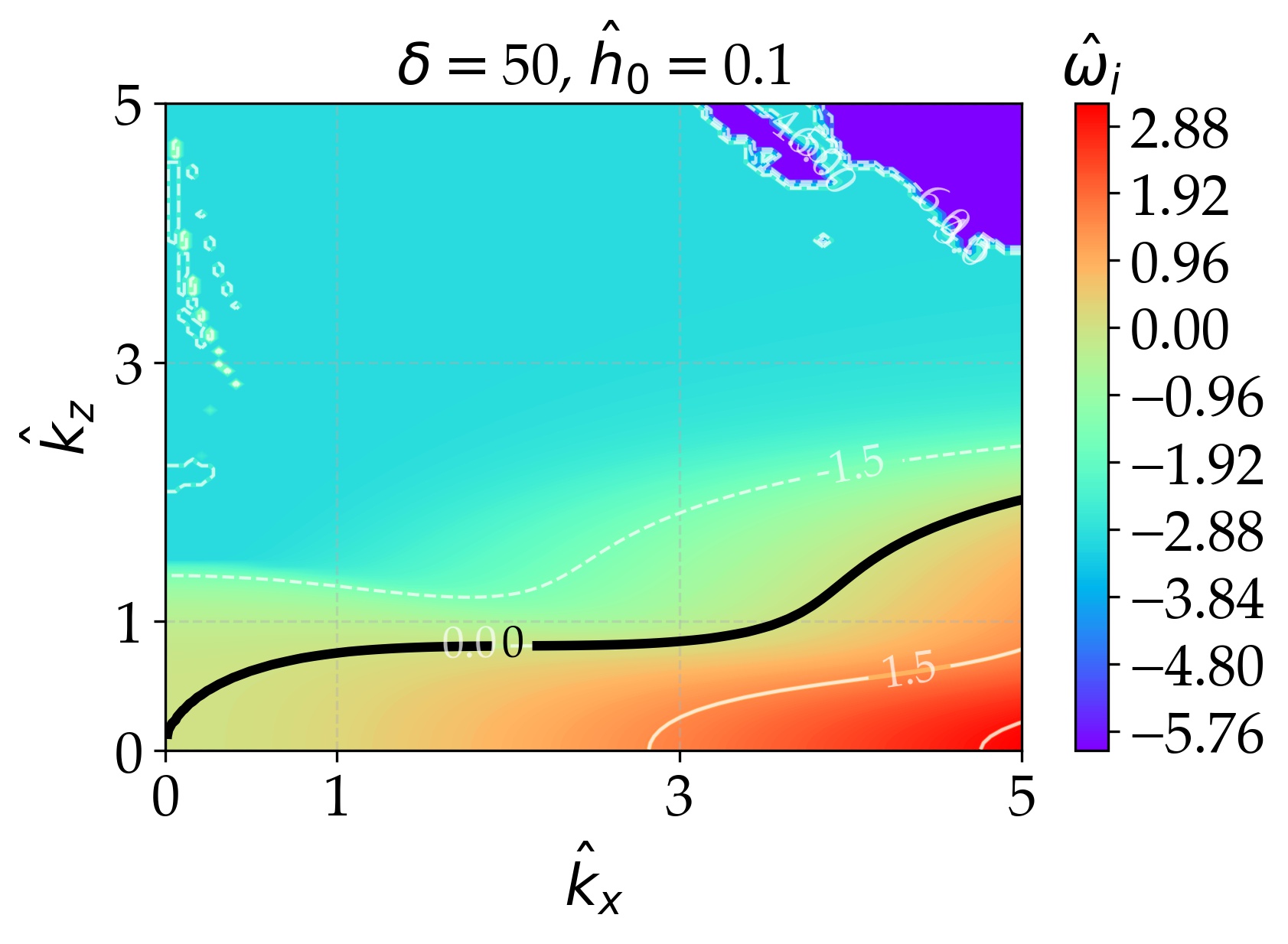}
  \caption{}
  \label{fig:h01_delta50}
\end{subfigure}\hfil % <-- added
\begin{subfigure}{0.33\textwidth}
  \includegraphics[width=\linewidth]{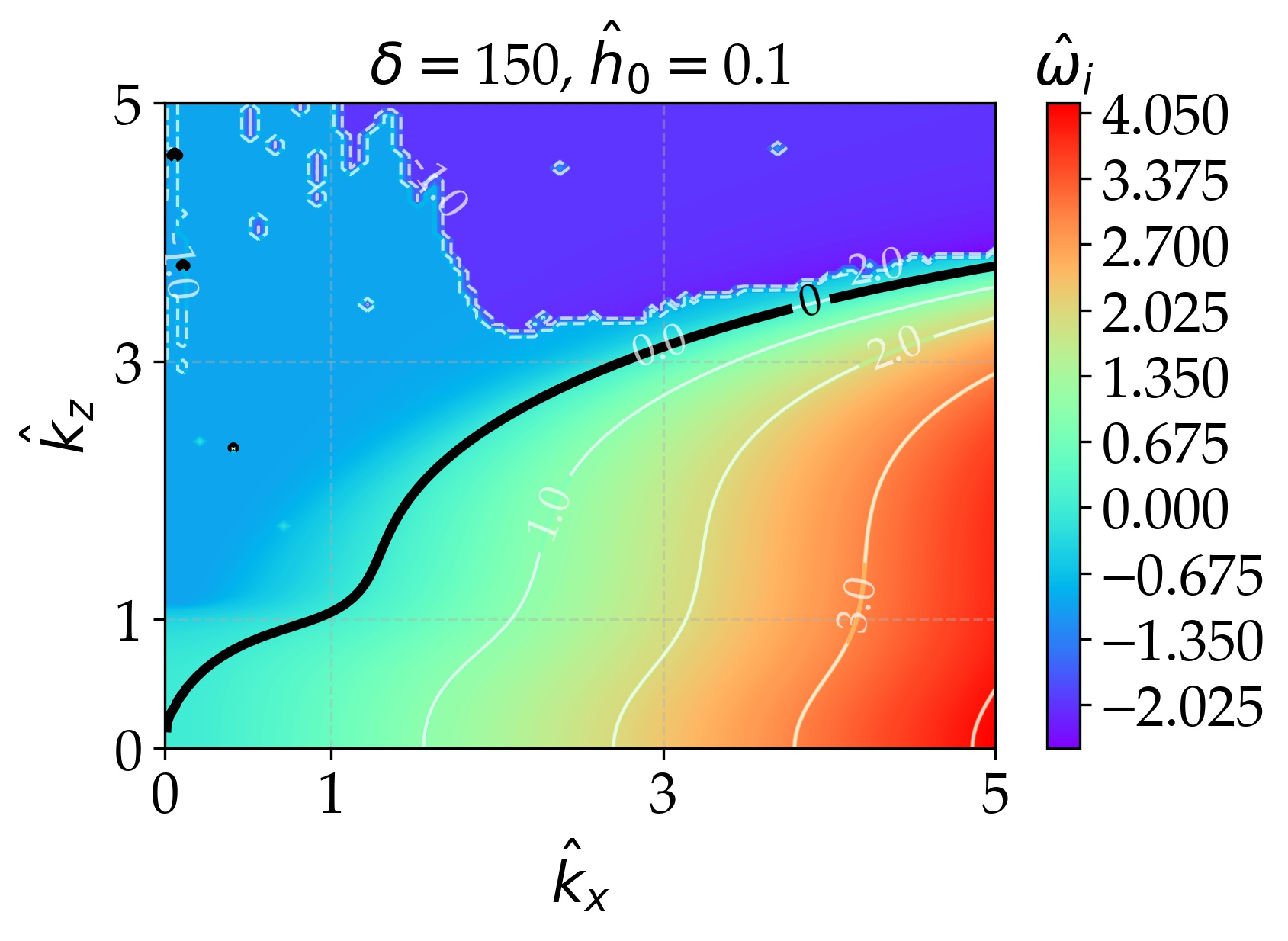}
  \caption{}
  \label{fig:h01_delta150}
\end{subfigure}\hfil % <-- added
\begin{subfigure}{0.33\textwidth}
  \includegraphics[width=\linewidth]{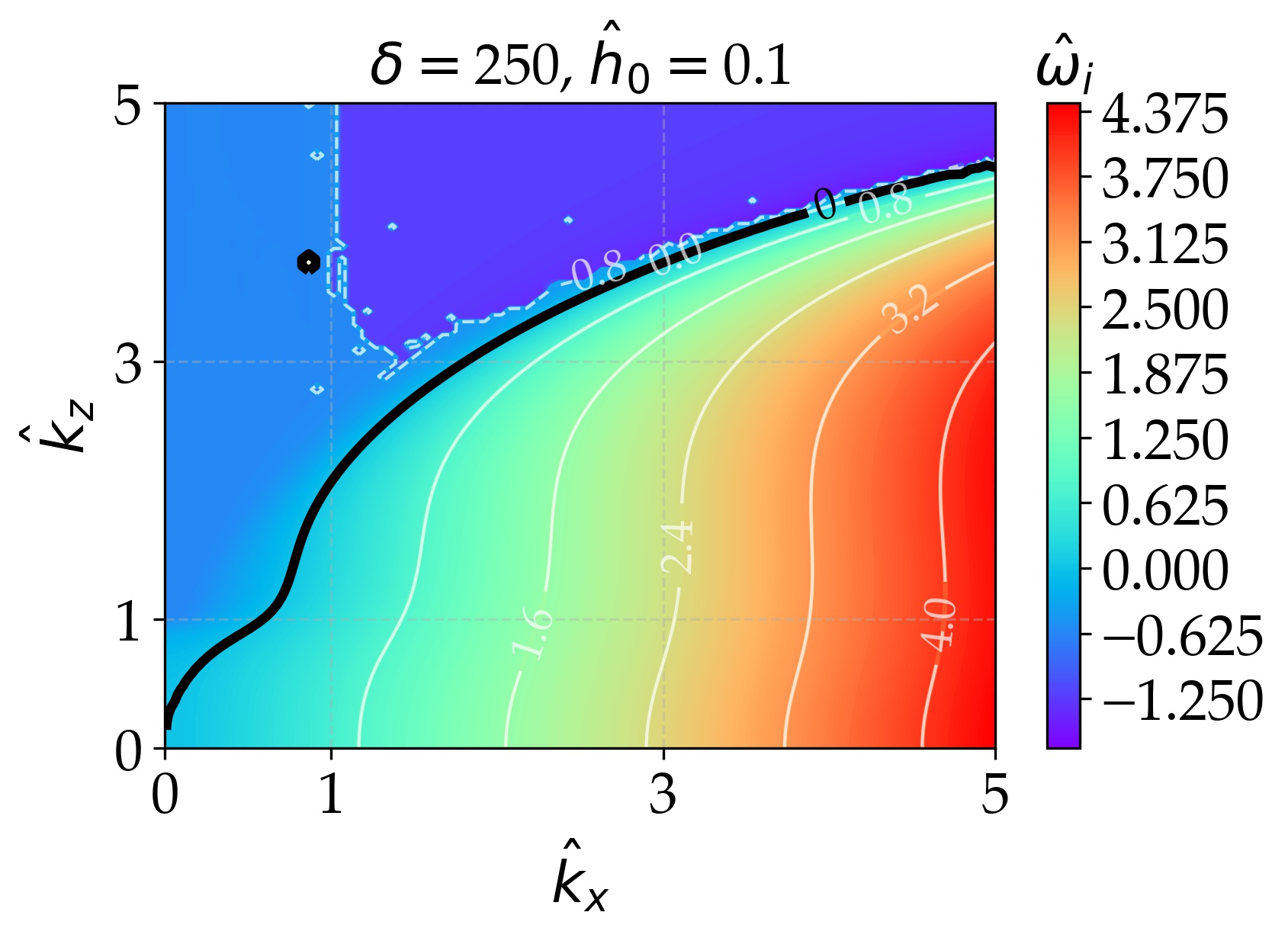}
  \caption{}
  \label{fig:h01_delta250}
\end{subfigure}

\medskip

\begin{subfigure}{0.33\textwidth}
  \includegraphics[width=\linewidth]{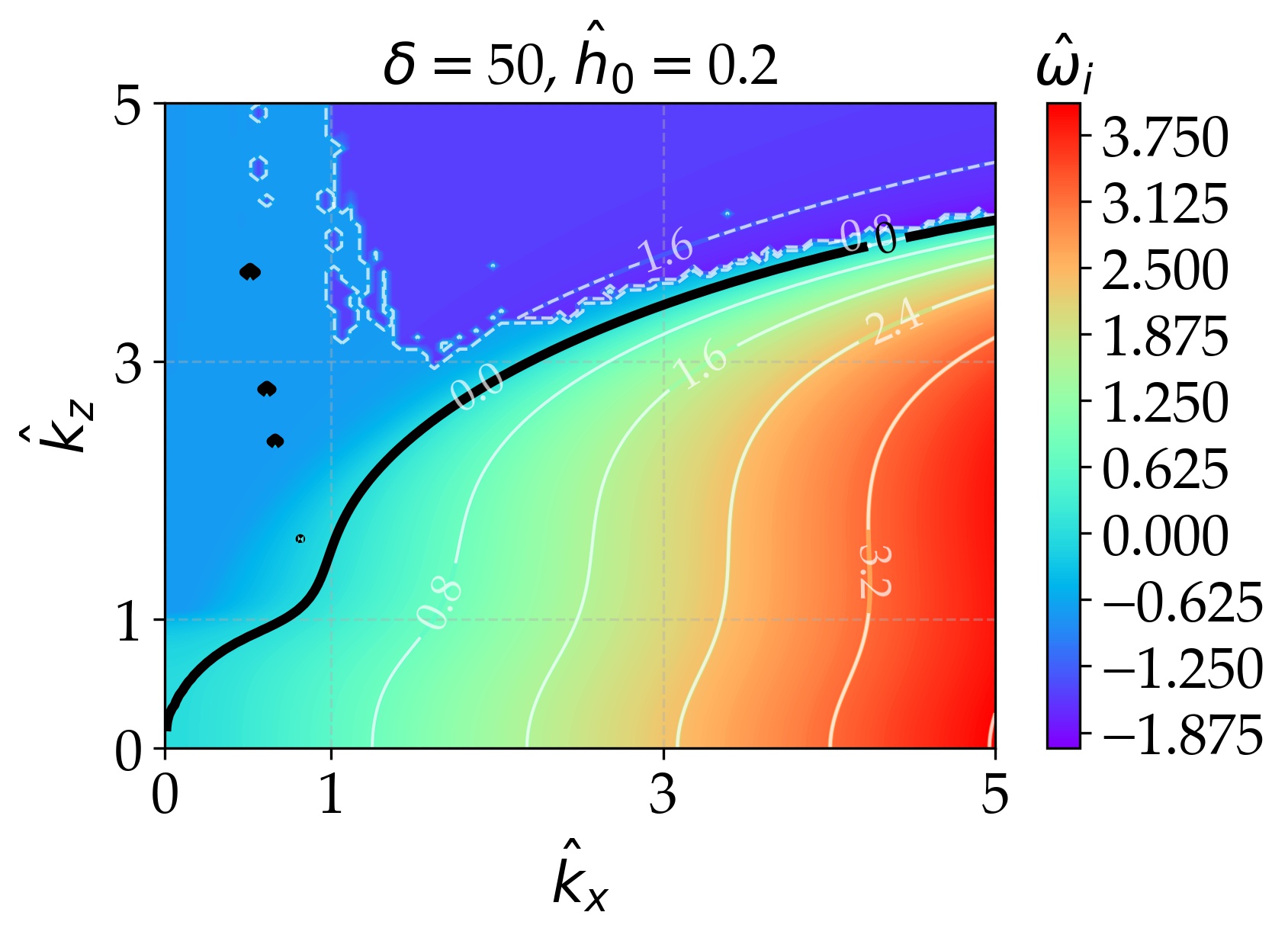}
  \caption{}
  \label{fig:h02_delta50}
\end{subfigure}\hfil % <-- added
\begin{subfigure}{0.33\textwidth}
  \includegraphics[width=\linewidth]{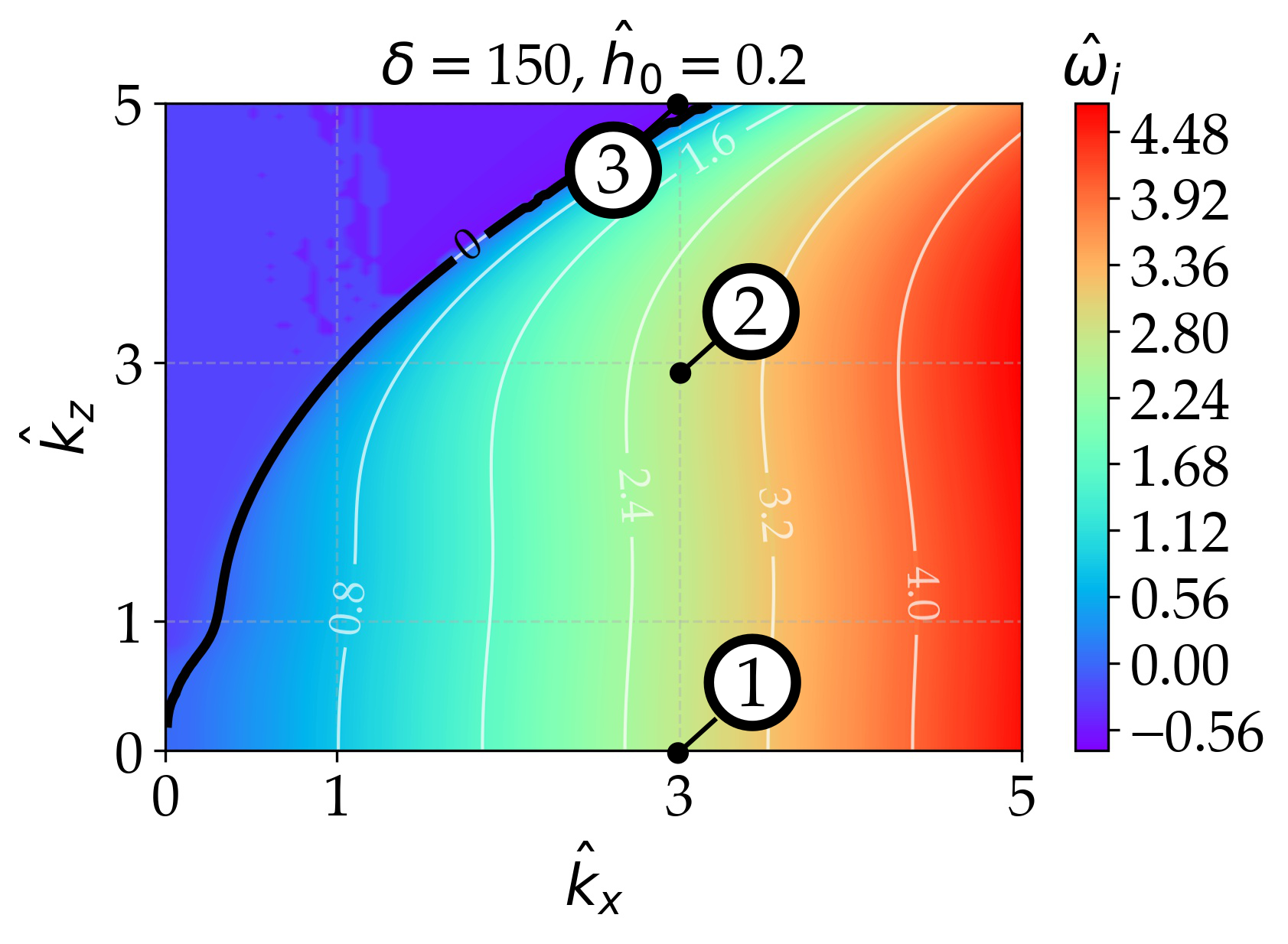}
  \caption{}
  \label{fig:h02_delta150}
\end{subfigure}\hfil % <-- added
\begin{subfigure}{0.33\textwidth}
  \includegraphics[width=\linewidth]{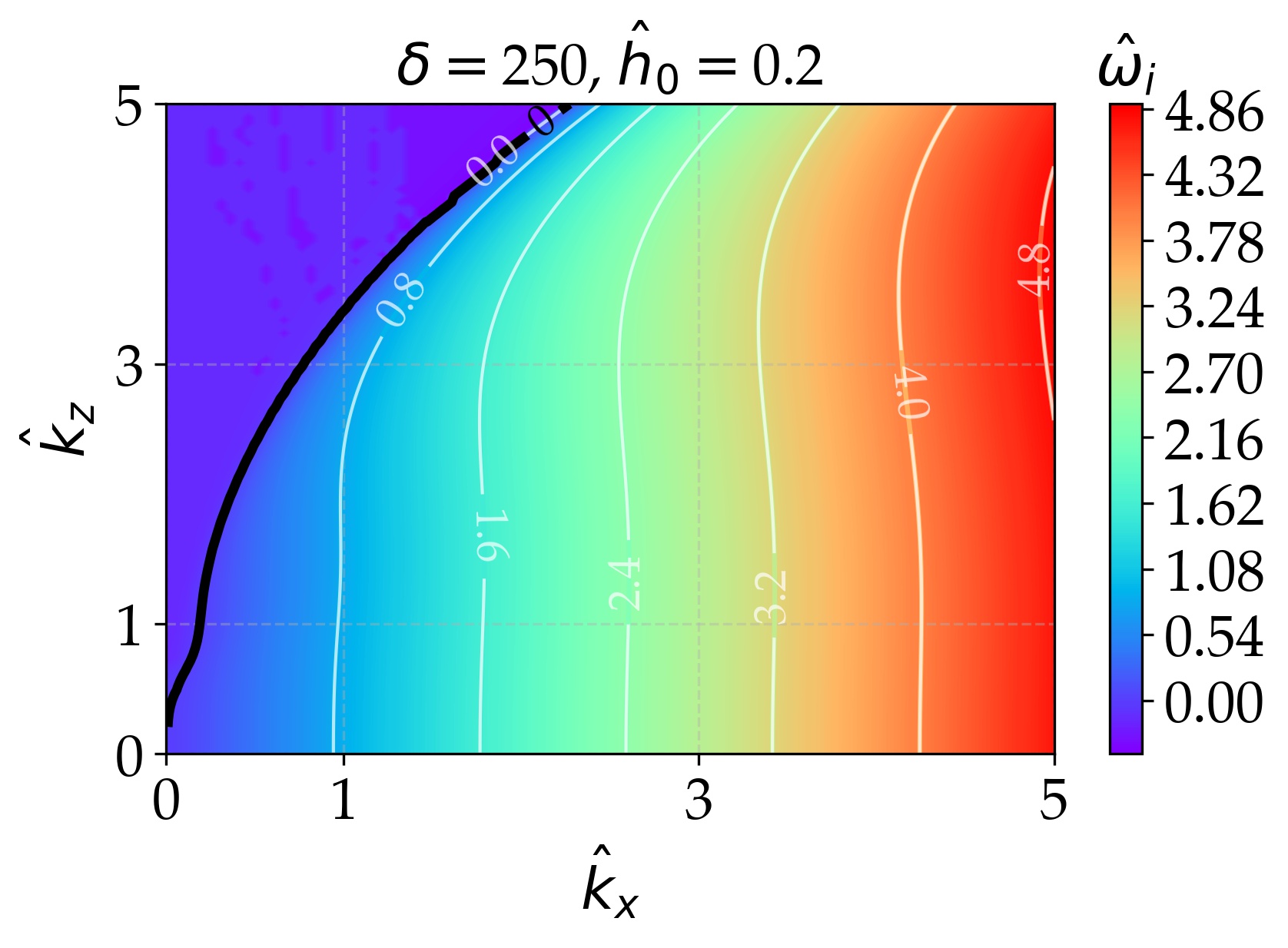}
  \caption{}
  \label{fig:h02_delta250}
\end{subfigure}
\caption{Colour map of $\omega_i$ for $\delta$=50, 150, and 250 for $\hat{h}_0$ = 0.05 (a,b,c), 0.1 (d,e,f) and 0.2 (g,h,i). The white lines show the isolines of $\omega_i$, and the black thick line corresponds to the neutral curve, i.e.: the boundary between the unstable \textcolor{black}{(below)} and stable \textcolor{black}{(above)} regions. The labels in Fig.~\ref{fig:h02_delta150} correspond to the test cases investigated in Section~\ref{sec:results_nonlinear}.}
\label{fig:linear_stability}
\end{figure*}

The results for the nine selected conditions are presented in Fig.~\ref{fig:linear_stability} in the form of contour plots of \textcolor{black}{the amplification factor} $\omega_i$. The black line separating the \textcolor{black}{linearly unstable} regions with $\omega_i>0$ \textcolor{black}{(below)} from those \textcolor{black}{linearly stable} with $\omega_i<0$ \textcolor{black}{(above)} indicates neutral stability. As expected, increasing $\hat{h}_0$ and $\delta$ reduces the stability region. For instance, when comparing the results at $\delta=150$ and different $\hat{h}_0$ (Fig.~\ref{fig:h005_delta150} and ~\ref{fig:h02_delta150}, respectively), it is found that the system becomes more unstable when $\hat{h}_0$ increases. The same occurs when $\hat{h}_0$ is fixed at $0.2$ and $\delta$ increases from $50$ to $250$ (Fig.~\ref{fig:h02_delta50} and ~\ref{fig:h02_delta250}, respectively). In addition, the ratio between the largest amplification and the largest attenuation factor is much higher in the most unstable case in figure~\ref{fig:h02_delta250} than in the most stable one in figure~\ref{fig:h005_delta50}. In the case $\hat{h}_0=0.05$ and $\delta = 50$, the largest amplification factor is null within two digits ($\omega_i\approx0$), while the strongest attenuation is characterized by $\omega_i=-23.4$. On the other hand, in the case $\hat{h}_0=0.2$ and $\delta=250$, the largest amplification is $\omega_i=4.86$, while the most stable conditions are close to neutral ($\omega_i\approx0$).

\textcolor{black}{Comparing these results to those of Ivanova et al.\cite{Ivanova2022_BLEW3D} for two-dimensional perturbations ($\hat{k}_z=0$), we observe that the system is globally more unstable when three-dimensional disturbances are considered. In the LSA, the stabilization of the system is solely due to the capillarity terms proportional to $\hat{k}^{3}$ on the RHS of \eqref{eq:stability_BLEW}; these act compensating the inertial destabilization generated by the terms proportional to $\hat{k}$ on the LHS. The interplay between these forces in both directions is also depicted in these stability maps. Inertia dominates over capillarity along the $\hat{k}_x$ axis, since the amplification factor $\omega_i$ is a growing function of $\hat{k}_x$ (the stability region at a given $\hat{k}_z$ is reduced at larger $k_x$), while the reverse is true along the $\hat{k}_z$ axis. The different behavior in each direction is due to the substrate motion, which might destabilize the flow in the stream-wise coordinate. On the other hand, increasing the $\hat{k}_z$ at constant $\hat{k}_x$ stabilizes the system because the capillary terms in \eqref{eq:stability_BLEW} grow faster than the inertial terms even at very small wave numbers.}

\begin{figure}
    \centering
	\includegraphics[width=0.8\linewidth]{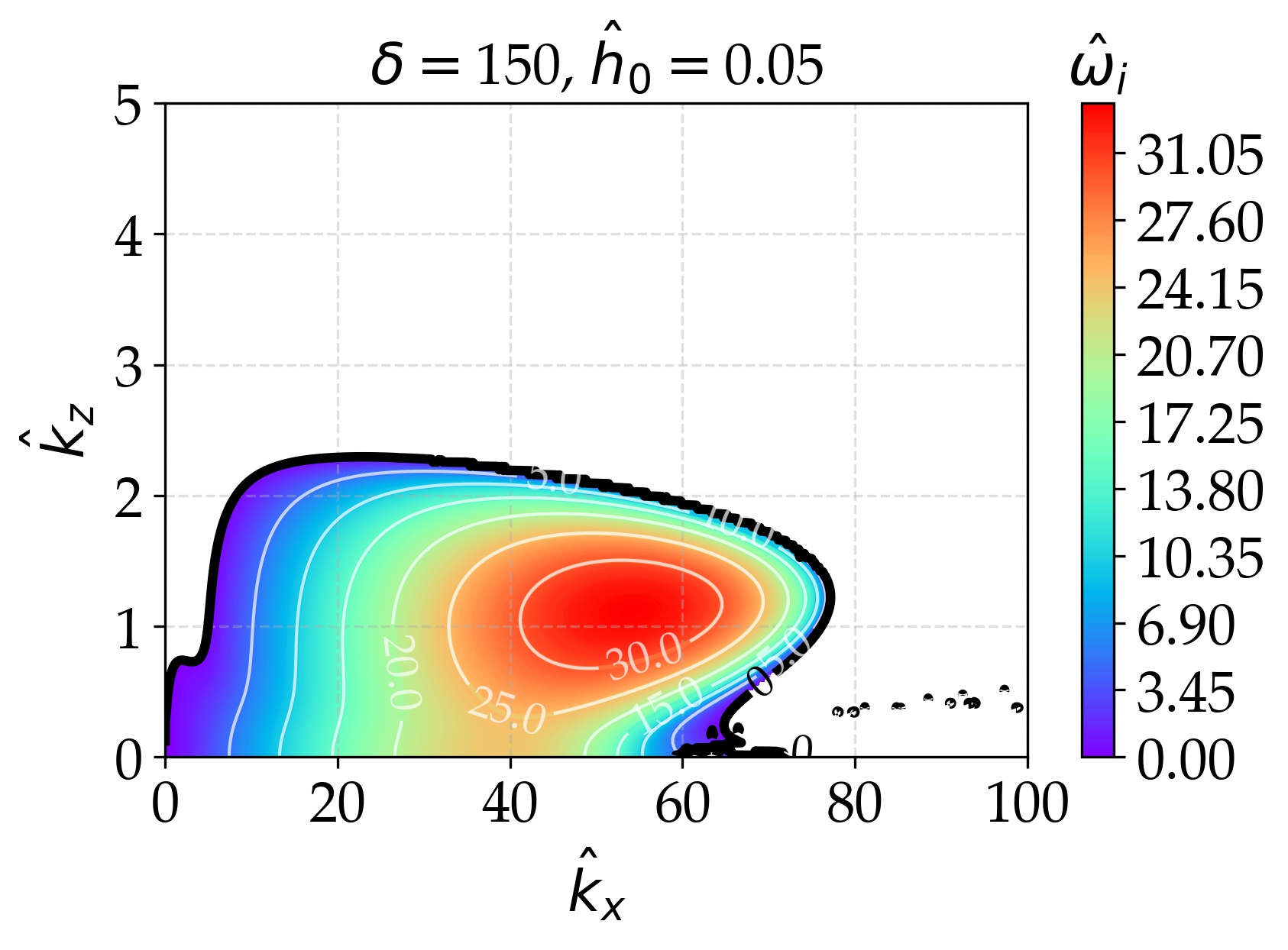} 
	\caption{\label{fig:global_LSA_delta150} Same plot as in Fig.~\ref{fig:linear_stability}, for the case with $\delta$=150 and $\hat{h}_0$ =0.05. The white area is the stable region ($\omega_i<0$). Above $\hat{k}_x > 80$ and $\hat{k}_x> 2.5$, all waves a linearly stable because of surface tension.} 
\end{figure}

\textcolor{black}{Following this rationale, the system must be linearly stable for sufficiently high wave numbers, above which capillarity always compensates inertia. This is shown in Fig.~\ref{fig:global_LSA_delta150} for $\hat{h}_0=0.05$ and $\delta=150$, which plots the stability map \textcolor{black}{over} a larger wave number space. The maximum amplification occurs at $\hat{\mathbf{k}}=(55,1)$, and surface tension restores the stability for all $ \hat{k}_z>2.5$. Nevertheless, these ranges correspond to non-physically high stream-wise wave numbers $\hat{k}_x$ in usual coating applications as shown in Table~\ref{tab:references_coating}, hence in the following sections, we focus on a narrower range of wave numbers, namely the ones shown in Fig.~\ref{fig:linear_stability}}.

\subsection{Validation of the film solver}
\label{sec:validation}

\begin{figure} \center
\begin{subfigure}{0.99\linewidth}
	\centering
	\includegraphics[width=\linewidth]{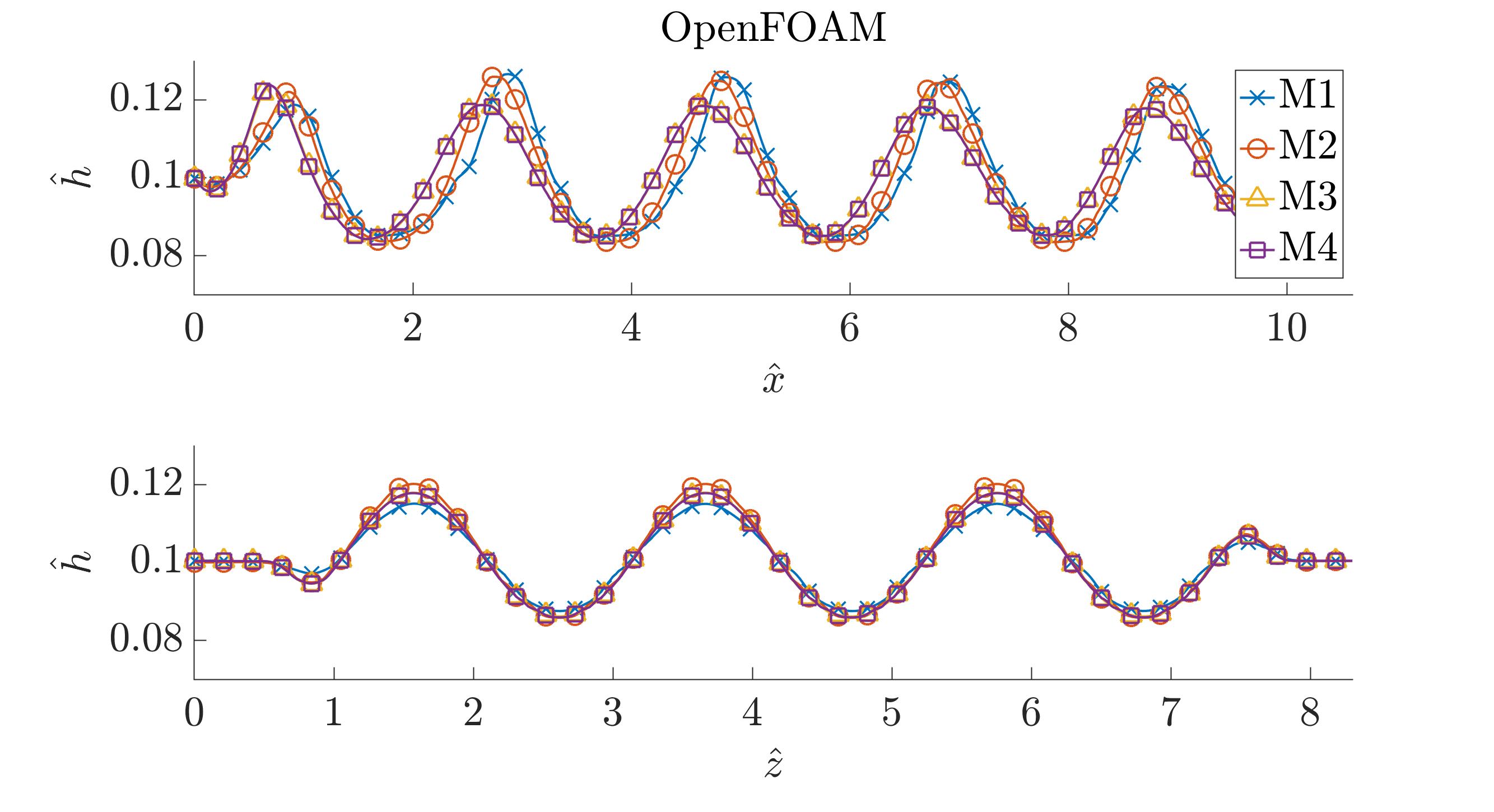}
	\caption{M1 (1.2 M cells): $n_\lambda=50$, $n_h=11$; M2 (8.8 M cells): $n_\lambda=100$, $n_h=22$; M3 (22.4 M cells): $n_\lambda=125$, $n_h=40$;  M4 (35.3 M cells): $n_\lambda=150$, $n_h=45$.}
	\label{fig:OF_mesh_sensitivity}
\end{subfigure}
\begin{subfigure}{0.99\linewidth}
	\centering
	\includegraphics[width=\linewidth]{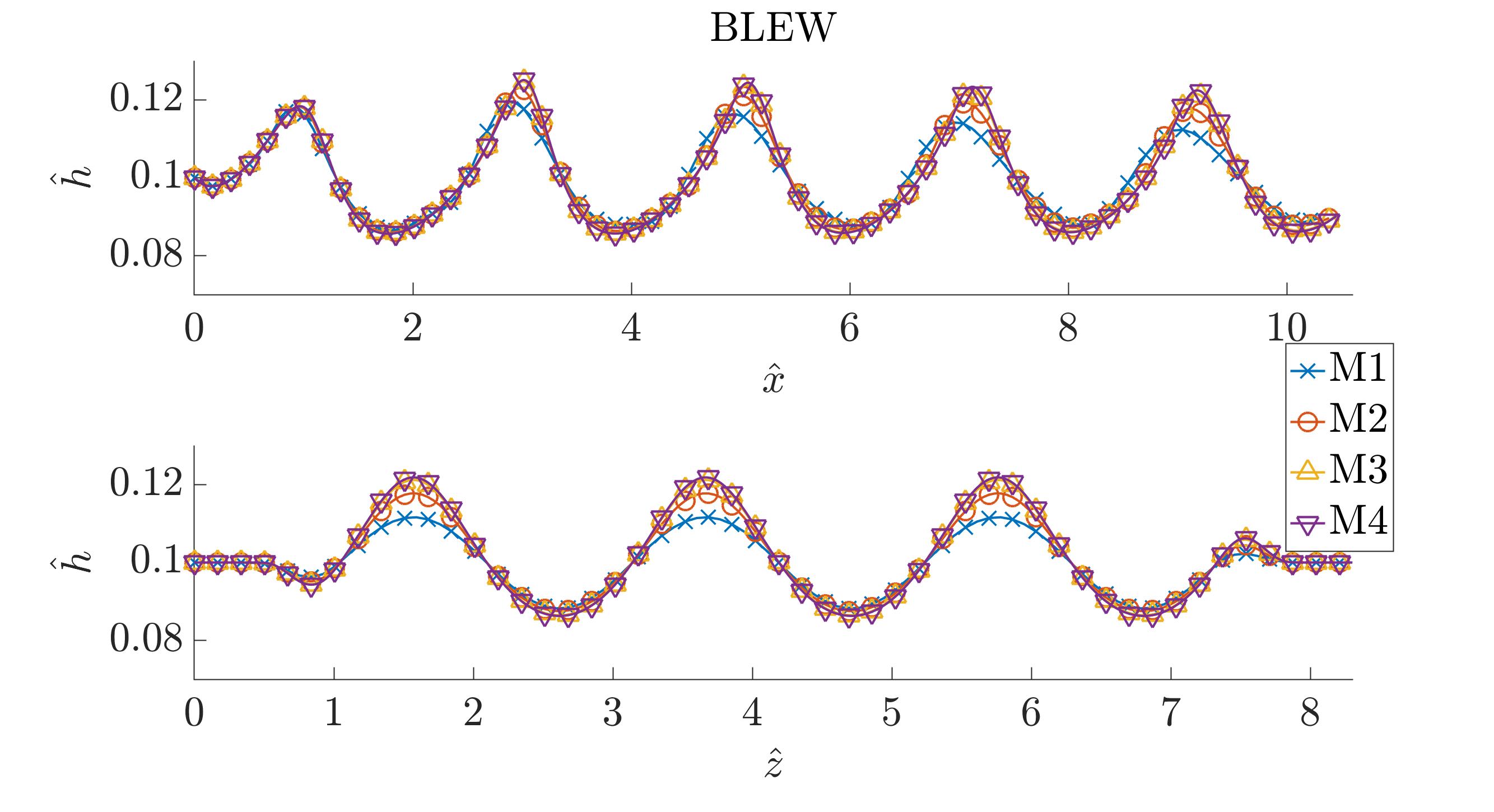}
	\caption{M1 (12.5 k cells): $n_\lambda=25$, M2 (50 k cells): $n_\lambda=50$ , M3 (200 k cells): $n_\lambda=100$, M4 (312.5 k cells): $n_\lambda=125$.}
	\label{fig:BLEW_mesh_sensitivity}
\end{subfigure}
\caption{Sensitivity analysis with respect to the mesh in OpenFOAM (a) and BLEW (b).}
\label{fig:mesh_sensitivity}
\end{figure}

The objective of this section is two-fold: (1) \textcolor{black}{report} on the sensitivity of the IBL and the DNS simulations to the numerical setup, and (2) \textcolor{black}{validate} the IBL simulations against DNS. \textcolor{black}{The working fluids are air and water with} fluid properties as detailed in Section~\ref{sec:OF_setup}, and substrate velocity $U_p$ equal to 1.45 m/s \textcolor{black}{in the direction against gravity}. The resulting Reynolds number is $Re = 558$ and the film parameter $\varepsilon=0.27$\textcolor{black}{, leading} to a reduced Reynolds number of $\delta=151$. \textcolor{black}{Once again, we stress that this is well beyond the expected limit of validity for the IBL model (Section \ref{sec:BLEW_Equations})}. The film thickness as base state is set to $\hat{h}_0 = 0.1$, and the stream-wise flow rate is perturbed as in \eqref{eq:3Dperturbation} with a relative amplitude $A=0.1$, frequency $\hat{f} = 0.48$, and span-wise wavelength $\hat{\lambda}_z = 1/ \hat{f}$ such that the wavelengths in both directions are identical. In dimensional form, this configuration corresponds to a mean film thickness of approximately 38 microns and wavelengths of 3 millimeters.

\begin{figure*}[ht!]
      \begin{subfigure}{0.99\linewidth}
    	\centering
    	\includegraphics[width=0.9\textwidth]{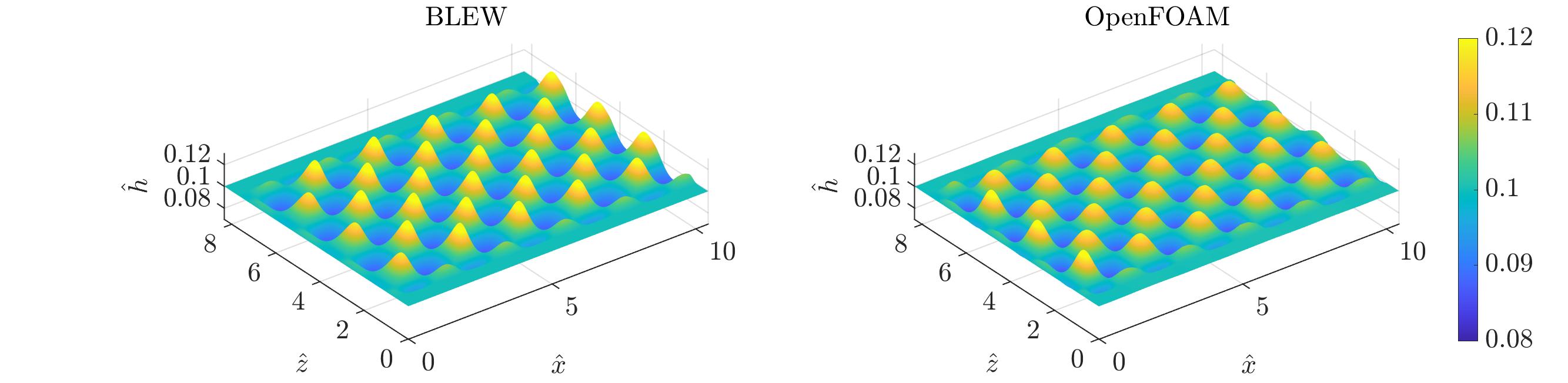} % Here is how to import EPS art
    	\caption{} \label{fig:3Dwaves_noST_validation}
         \end{subfigure}
         
      \begin{subfigure}{0.9\linewidth}
    	\centering
    	\includegraphics[width=\linewidth]{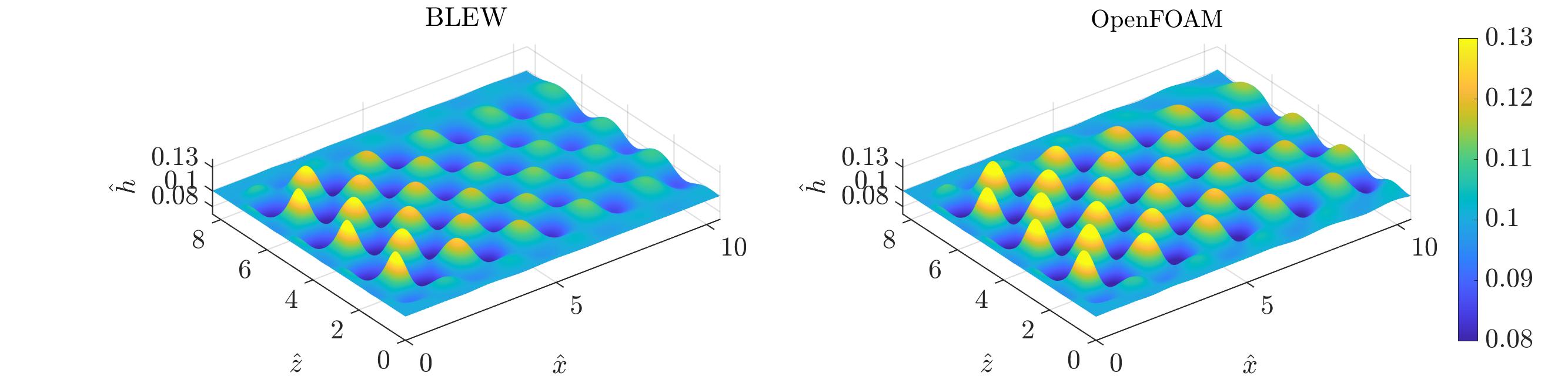} % Here is how to import EPS art
    	\caption{} \label{fig:3Dwaves_ST_validation}
         \end{subfigure}
        \caption{Snapshots of the three-dimensional evolution of the waves. Comparison of the IBL simulation in BLEW (left) and DNS in OpenFOAM (right). The cases in (a) are computed without surface tension; cases in (b) have surface tension active. \textcolor{black}{All snapshots are taken at a sufficiently large time, to let the wave pattern establish.}}
    \end{figure*}

We start by analyzing the sensitivity of the computations in  absence of surface tension, \textcolor{black}{in order} to validate the pseudo-spectral implementation of the surface tension terms in an independent way. Fig.\ref{fig:mesh_sensitivity} \textcolor{black}{documents} the mesh sensitivity of the DNS simulations (a) and IBL simulations (b) through the instantaneous thickness profiles along $\hat{x}$ and $\hat{z}$. The top plot shows the downstream evolution of waves in a longitudinal plane at $\hat{z}= \hat{\lambda_z}/2 + \hat{\lambda_z}/4$ (where the perturbation amplitude is maximum)\textcolor{black}{, and} the bottom one depicts the transverse profiles \textcolor{black}{taken} at the crest of the wave closest to the outlet. Note that the y-axis --of the order of micrometers-- is significantly enlarged compared to $x$ and $z$ --of the order of millimeters-- for plotting purposes in all figures.

The DNS has an additional degree of freedom because the cross-stream direction is also  discretized. Therefore, the number of cells per wavelength $n_\lambda$ is varied together with the cell number across the film, $n_h$, as detailed in the caption of Fig.~\ref{fig:OF_mesh_sensitivity}. The DNS was found to be more dependent on the mesh resolution, where a minimum cell size in both stream/span-wise and cross-stream directions is required to fully capture the physics of the problem. The mesh M3, consisting of 125 cells per wavelength and 40 cells across the film (22.4 M cells in total) was sufficient to reach mesh independence in OpenFOAM. Similarly, the minimum number of cells per wavelength $n_\lambda$ is approximately 100 in the IBL simulations. %(It is interesting to note that coarse meshes in OpenFOAM lead to higher amplitudes, while the opposite occurs in BLEW. This is a clear consequence of the different paradigms behind both approaches. (I would remove that))

% DAVID: General remark: emphasize DNS vs IBL rather than OpenFoam versus BLEW.

\begin{figure}[h]
    \centering     
    	\includegraphics[width=\linewidth]{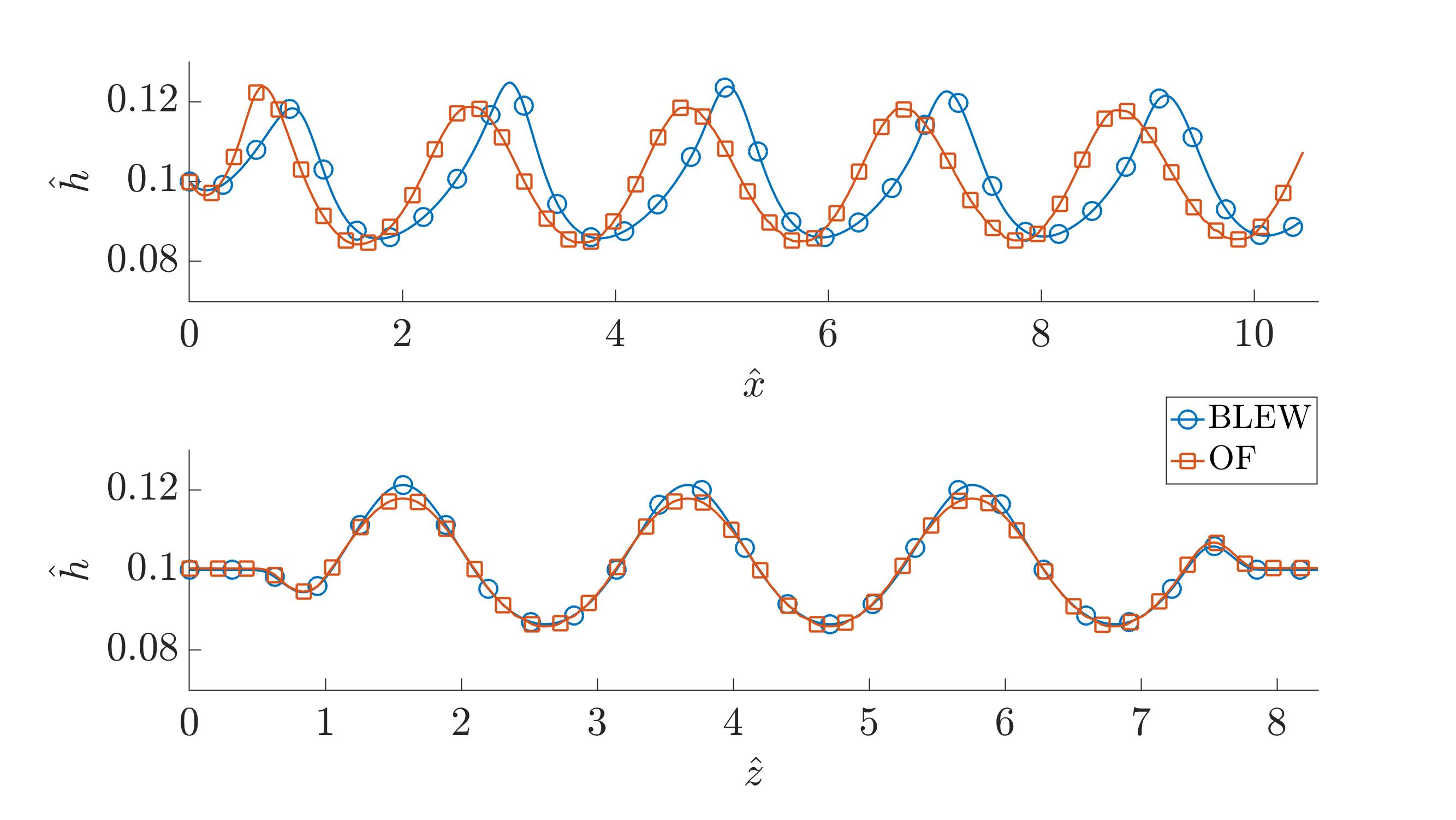} % Here is how to import EPS art
    	\caption{Comparison between the IBL simulations (in BLEW) and the DNS simulations (in OpenFOAM, OF) in the absence of surface tension. The stream-wise ($x$) thickness evolution is displayed on the top, and the span-wise (z) on the bottom.} 
\label{fig:OF_BLEW_noST_validation}        
\end{figure}

The mesh independent solutions are qualitatively confronted in Fig.~\ref{fig:3Dwaves_noST_validation} with instantaneous three-dimensional thickness distributions and in 
Fig.~\ref{fig:OF_BLEW_noST_validation} with longitudinal and transversal thickness profiles. Comparing the stream-wise profiles, the initial growth of the perturbation is slightly different: the wave closer to the inlet in the DNS gets steeper and higher compared to IBL. This is probably due to the differences in the inlet boundary conditions (i.e. perturbation of the flow rate in IBL, and velocity in DNS) and modeling strategies (i.e. the interaction between the liquid film and the quiescent atmosphere is not considered in the film solver). However, the agreement is satisfactory in the sense that the shape, amplitude, and speed of the waves remain rather unaltered in both cases. This is further supported by the excellent matching observed in the transverse profiles in Fig.~\ref{fig:OF_BLEW_noST_validation}. 

\begin{figure}
    \centering

     \begin{subfigure}{0.99\linewidth}
    	\centering
    	\includegraphics[width=\linewidth]{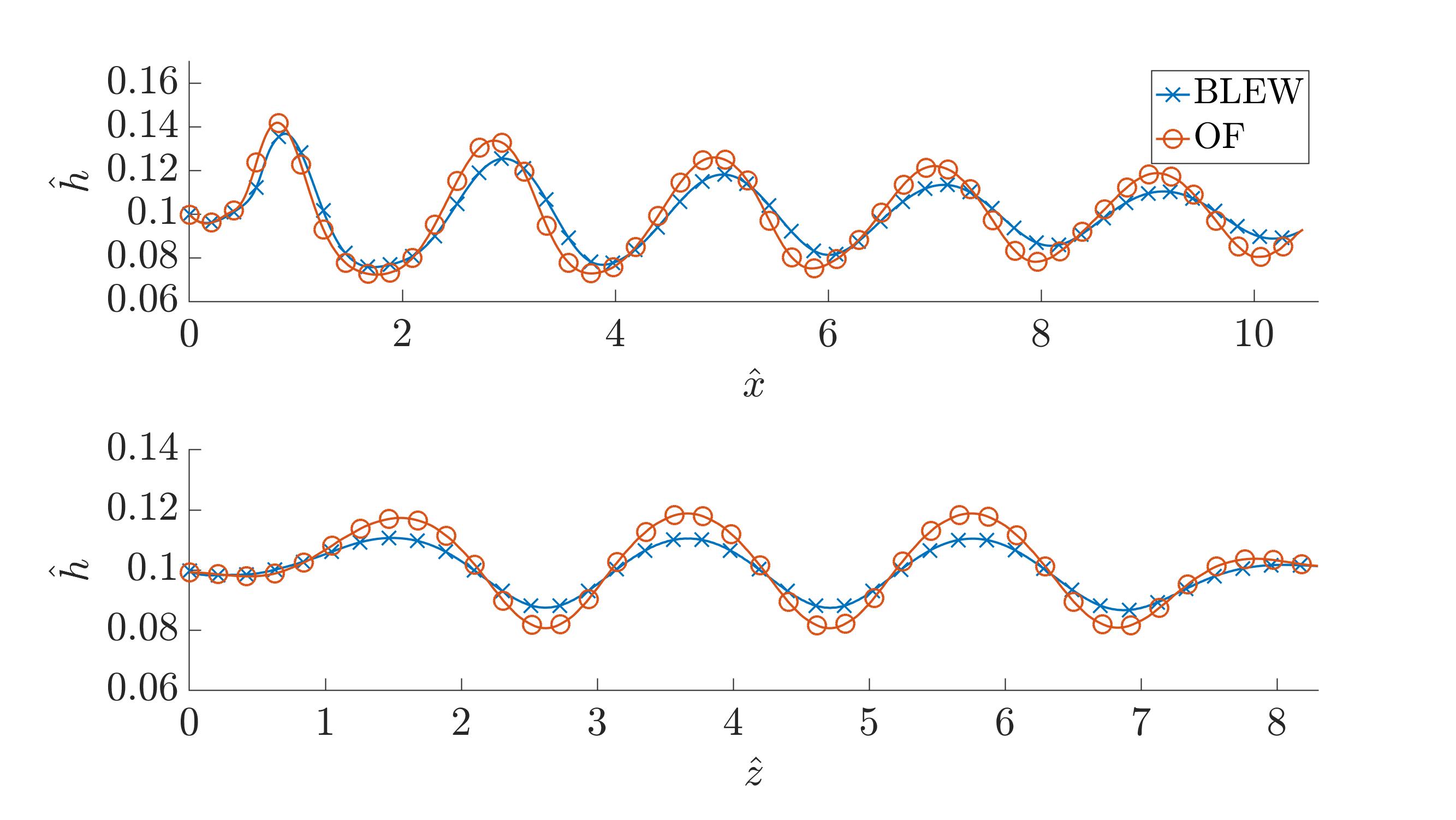} % Here is how to import EPS art
    	\caption{} \label{fig:2D_profiles_ST_validation}
    \end{subfigure}
    
     \begin{subfigure}{0.99\linewidth}
    	\centering
    	\includegraphics[width=\linewidth]{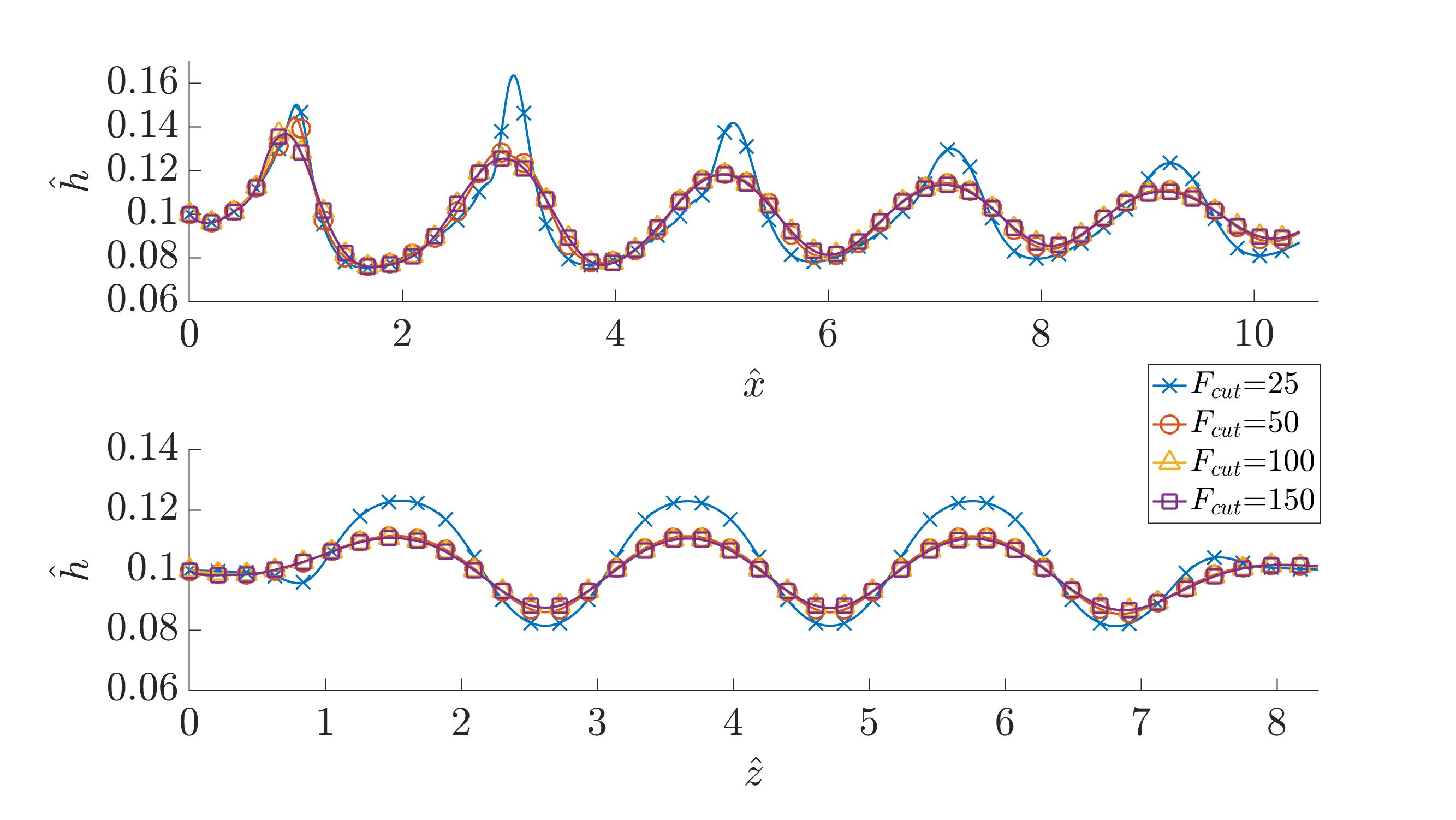} % Here is how to import EPS art
    	\caption{} \label{fig:2D_profiles_ST_dependency}
    \end{subfigure}
\caption{\label{fig:OF_BLEW_ST_validation} Comparison \textcolor{black}{in terms thickness profiles along $x$ (top) and $z$ (bottom) predicted by IBL and DNS (a), and sensitivity of the results to the cut-off parameter $F_{cut}$ of the Butterworth filter (b). The test case is the one in Fig.~\ref{fig:OF_BLEW_noST_validation} but with larger perturbation ($A=0.2$) and including the surface tension.}}
\end{figure}

%The same validation is repeated in presence of surface tension and increasing the perturbation amplitude to $A=0.2$ \textcolor{black}{to enhance capillary forces}. 
The same validation is repeated \textcolor{black}{with the surface tension terms in Eq.~\ref{eq:BLEW_EQ},} and increasing the perturbation amplitude to $A=0.2$ \textcolor{black}{to enhance capillary forces}. \textcolor{black}{This case aims at validating the pseudo-spectral implementation for the surface tension introduced in Section~\ref{sec:BLEW_Equations} in capillary dominated film flows. The comparison between IBL and DNS simulations is illustrated with two instantaneous thickness distributions in Fig.~\ref{fig:3Dwaves_ST_validation} and thickness profiles in Fig.~\ref{fig:OF_BLEW_ST_validation}.} The \textcolor{black}{3D} plot shows that the qualitative agreement is satisfactory since both solvers capture the initial growth of the waves, followed by capillary damping. The differences are more evident in the profiles in Fig.~\ref{fig:2D_profiles_ST_validation}; however, the agreement is \textcolor{black}{fair} taking into account the approximations of the IBL model and the uncertainty of the VOF computations \textcolor{black}{due to the diffusion at the interface}. The delay of the DNS signal compared to the IBL previously observed in Fig.~\ref{fig:OF_BLEW_noST_validation} seems to be mitigated when surface tension is active since it limits the abrupt wave growth at the inlet. Finally, the impact of the filtering of the capillary terms is shown in Fig.~\ref{fig:2D_profiles_ST_dependency}, which compares the downstream evolution of the waves for several cut-off parameters $F_{cut}$ ranging from 25 to 150. Setting this parameter between 50 and 150 is found to be sufficient to get independent results, which proves that the filtering does not affect the scales associated with the perturbations. \textcolor{black}{On the other hand, the simulations with $F_{cut}>175$ diverge due to noise amplification}. These results confirm that our proposed pseudo-spectral implementation is suitable to preserve the numerical stability of the solver without compromising the accuracy of the results. \textcolor{black}{More remarkably, these results validate the predictions of the IBL model in conditions that are well outside the asymptotic limits of its derivation.}

Finally, it is worth stressing that the computational cost of DNS is orders of magnitude larger than that of the IBL simulations due to the mesh requirements and the complexity of the governing equations. To give an order of magnitude, simulating one flow-through in DNS requires approximately 12 hours on 64 Intel E5-2680v3 CPUs at the Centro de Supercomputación de Galicia (CESGA), while the equivalent in IBL simulations requires 5 minutes on a single core of the same machine.

\subsection{Role of capillarity and non-linearities on wave damping}
\label{sec:results_nonlinear}

We now focus on the effect of capillary and non-linear terms on the damping of waves using both the IBL and the DNS. The results are also confronted with the LSA in Section~\ref{sec:lin_stability}. The aim was to assess the validity of these tools in the decision-making process in coating lines.

\begin{figure*}[htb]
    \centering % <-- added
    \begin{subfigure}{\textwidth}
      \includegraphics[width=\linewidth]{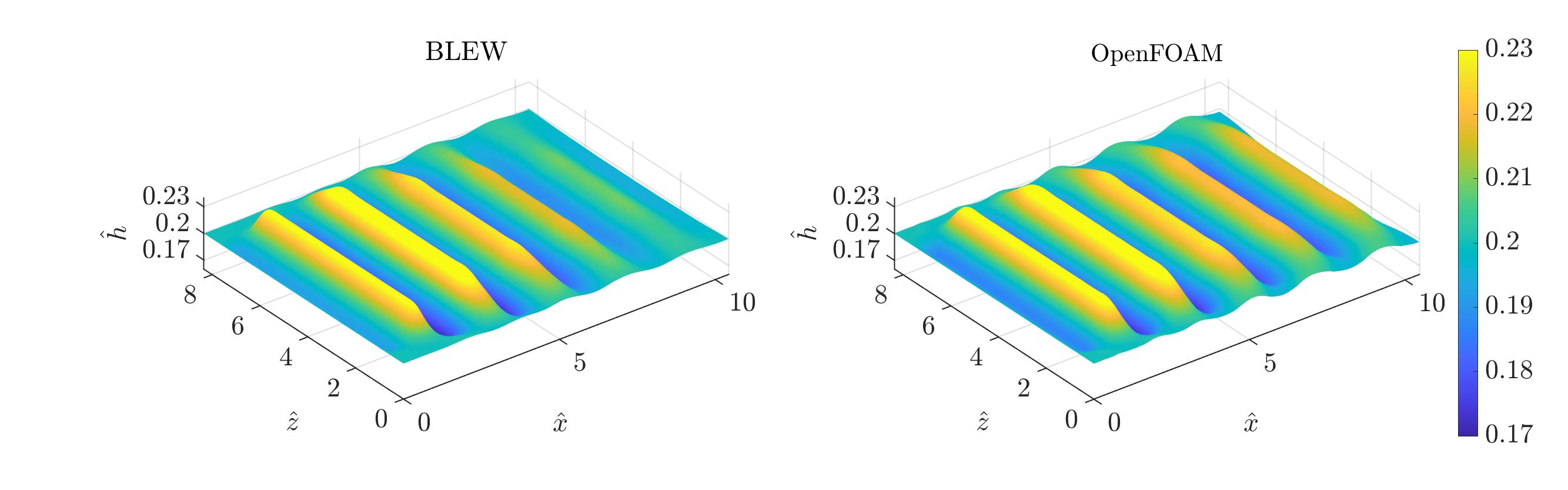}
      \caption{\textcolor{black}{Case 1:} $\hat{k}_x = 3$ , $\hat{k}_z = 0$ }
      \label{fig:kx3_2D}
    \end{subfigure}\hfil % <-- added

    \medskip
    
    \begin{subfigure}{\textwidth}
      \includegraphics[width=\linewidth]{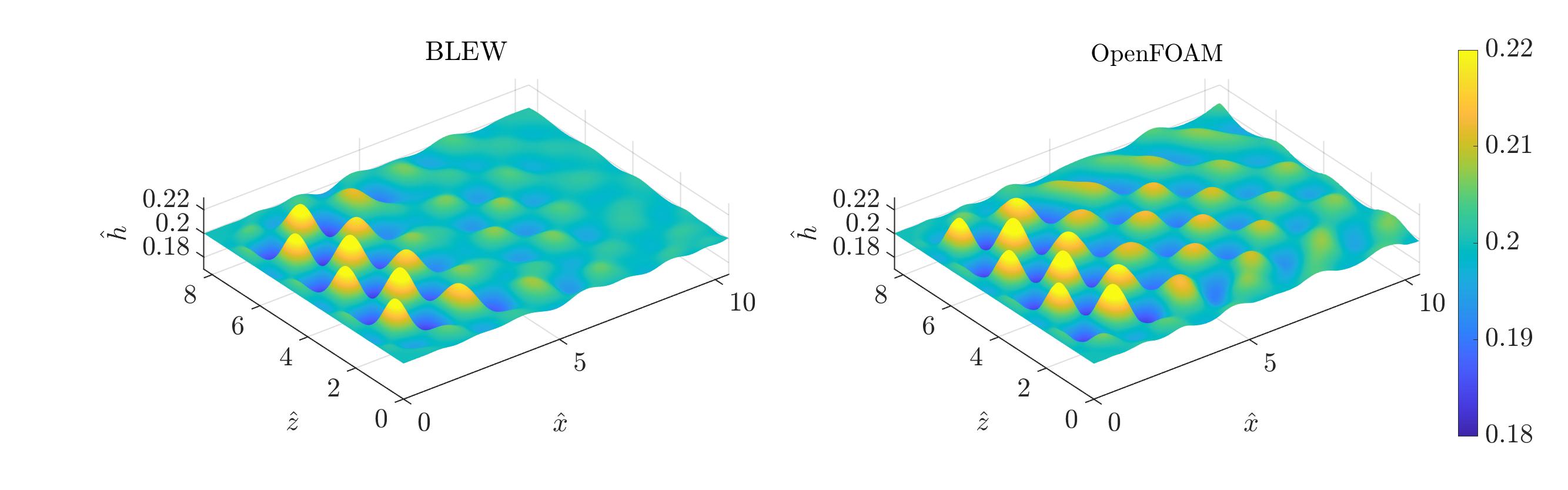}
      \caption{\textcolor{black}{Case 2:} $\hat{k}_x = 3$ , $\hat{k}_z = 3$ }
      \label{fig:kx3_kz3}
    \end{subfigure}\hfil % <-- added
    
    \medskip
    
    \begin{subfigure}{\textwidth}
      \includegraphics[width=\linewidth]{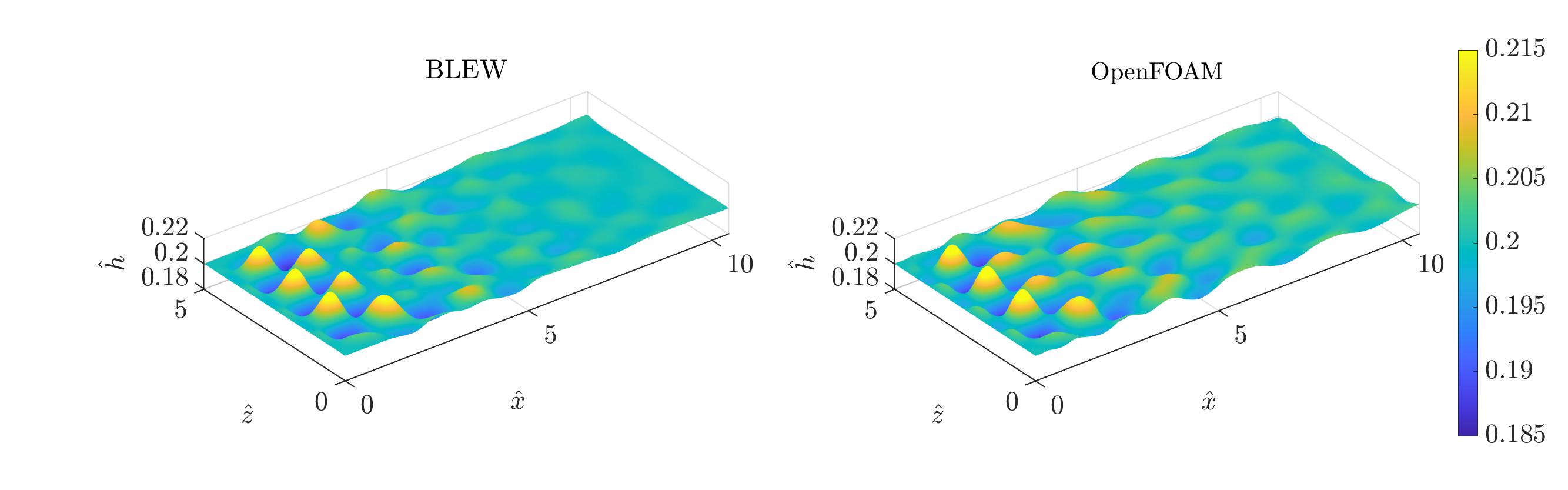}
      \caption{\textcolor{black}{Case 3:} $\hat{k}_x = 3$ , $\hat{k}_z = 5$ }
      \label{fig:kx3_kz5}
    \end{subfigure}\hfil % <-- added

\caption{\label{fig:LSA_vs_BLEW_vs_OpenFOAM} Three-dimensional evolution of the film thickness for \textcolor{black}{case 1 (a), 2 (b) and 3 (c)} with the stream and span-wise wave numbers indicated in the captions. The results obtained with the IBL solver BLEW are shown on the left, and the DNS OpenFOAM computations on the right.}
\end{figure*}

We consider the same test case used in the previous section for validation ($U_p=1.45 m/s$, $\delta=150$, $\epsilon = 0.27$), although the water film thickness is now set at $\hat{h}_0 = 0.2$ ($\approx 76 \mu m$ in the investigated conditions) and the perturbation amplitude is $A=0.05$. At this stage, we opted for a rather thick film to have large effects of the perturbation amplification and make the capillary effect more evident.

The mapping of linear amplification $\hat{\omega}_i$ for this configuration is illustrated in Fig.~\ref{fig:h02_delta150}; the labels identify three different flow configurations with identical stream-wise wave number $\hat{k}_x = 3$ and three span-wise wave numbers: label (1) corresponds to a 2D disturbances ($\hat{k}_z = 0$); labels (2) and (3) correspond to 3D disturbances, ($\hat{k}_z = 3$ and $\hat{k}_z = 5$ respectively). We thus focus on the effect of span-wise disturbances on the film response. The qualitative evolution of the waves in these three cases is illustrated in Fig.~\ref{fig:LSA_vs_BLEW_vs_OpenFOAM}  (IBL on the left and DNS on the right). The agreement between the two numerical solvers (and models) is excellent. \textcolor{black} {Once again, this result is surprising and noteworthy, considering that the conditions are well outside the theoretical range of validity of the IBL model.} 

\begin{figure*}
    \centering
      \includegraphics[width=\linewidth]{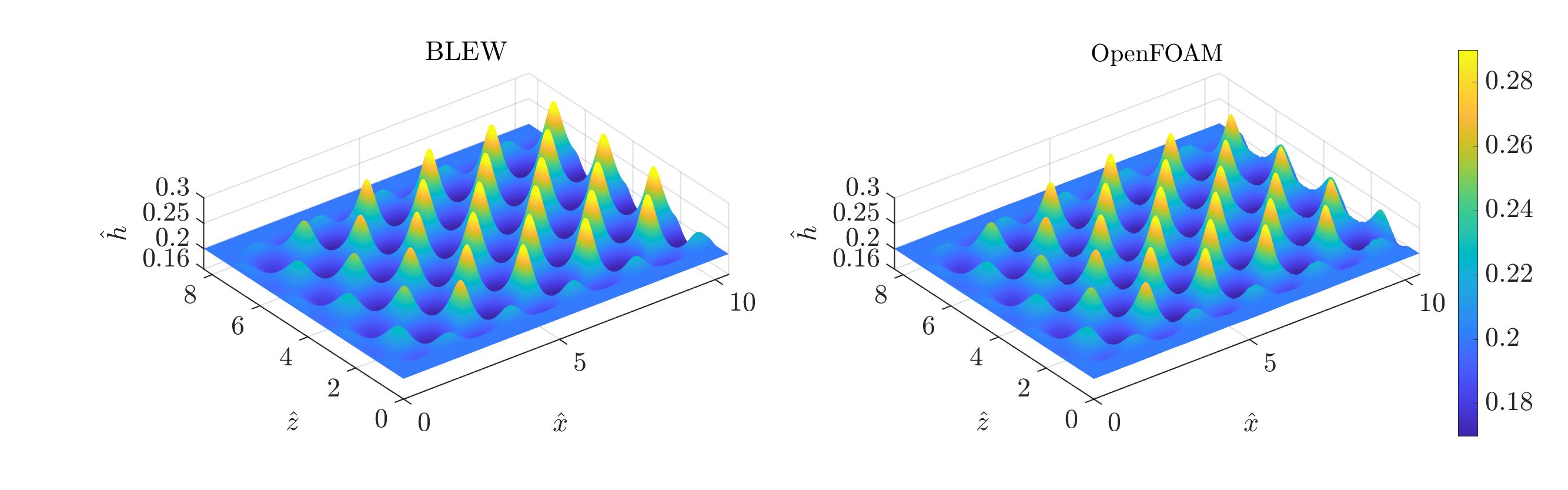}
      \caption{\label{fig:kx3_kz3_noST} Three-dimensional evolution of the film thickness for \textcolor{black}{case 2} in Fig.~\ref{fig:kx3_kz3} in absence of surface tension.}
\end{figure*}

Although the wave damping is evident, especially for three-dimensional disturbances, at this stage it is not possible to \textcolor{black}{determine} to which extent the damping is due to surface tension or to the non-linear stabilization mechanism identified in Ivanova et al.\cite{Ivanova2022_BLEW3D}. This is why we repeated the DNS computation of \textcolor{black}{case 2} ($\hat{k}_x = \hat{k}_z = 3$) \textcolor{black} setting zero surface tension. The resulting wave evolution in Fig.~\ref{fig:kx3_kz3_noST} shows a fast increase of the wave amplitude until saturation, where the amplitude reaches approximately 25 \% of the base state film thickness. These results confirm that for these conditions, the amplitude decay observed in Fig.~\ref{fig:LSA_vs_BLEW_vs_OpenFOAM} is entirely due to the effect of surface tension and not to the non-linear damping. The effect of the latter depends on the frequency and film thickness\cite{Ivanova2022_BLEW3D} and is further discussed with DNS computations at the end of this section.

Returning to the analysis of the results in Fig.~\ref{fig:LSA_vs_BLEW_vs_OpenFOAM}, in all the investigated cases, the film presents the usual abrupt initial growth followed by a progressive smoothing due to surface tension. This effect is much more evident in the 3D cases, as shown in Fig.~\ref{fig:kx3_kz3} for $\hat{k}_z = 3$ and Fig.~\ref{fig:kx3_kz5} for $\hat{k}_z = 5$. In these cases, the strong capillary-induced damping is dominated by the 3D-induced effects of surface tension. This results in the vanishing of waves after one of two wavelengths from their full growth. This effect, which becomes increasingly dominant as the span-wise wave number increases, can be well appreciated in the structure of the IBL model in \eqref{eq:BLEW_EQ}. The \textit{simplicity} of the governing equations allows for analyzing the relative contribution of the terms in various conditions.

\begin{figure}
    \centering
    \includegraphics[width=\linewidth]{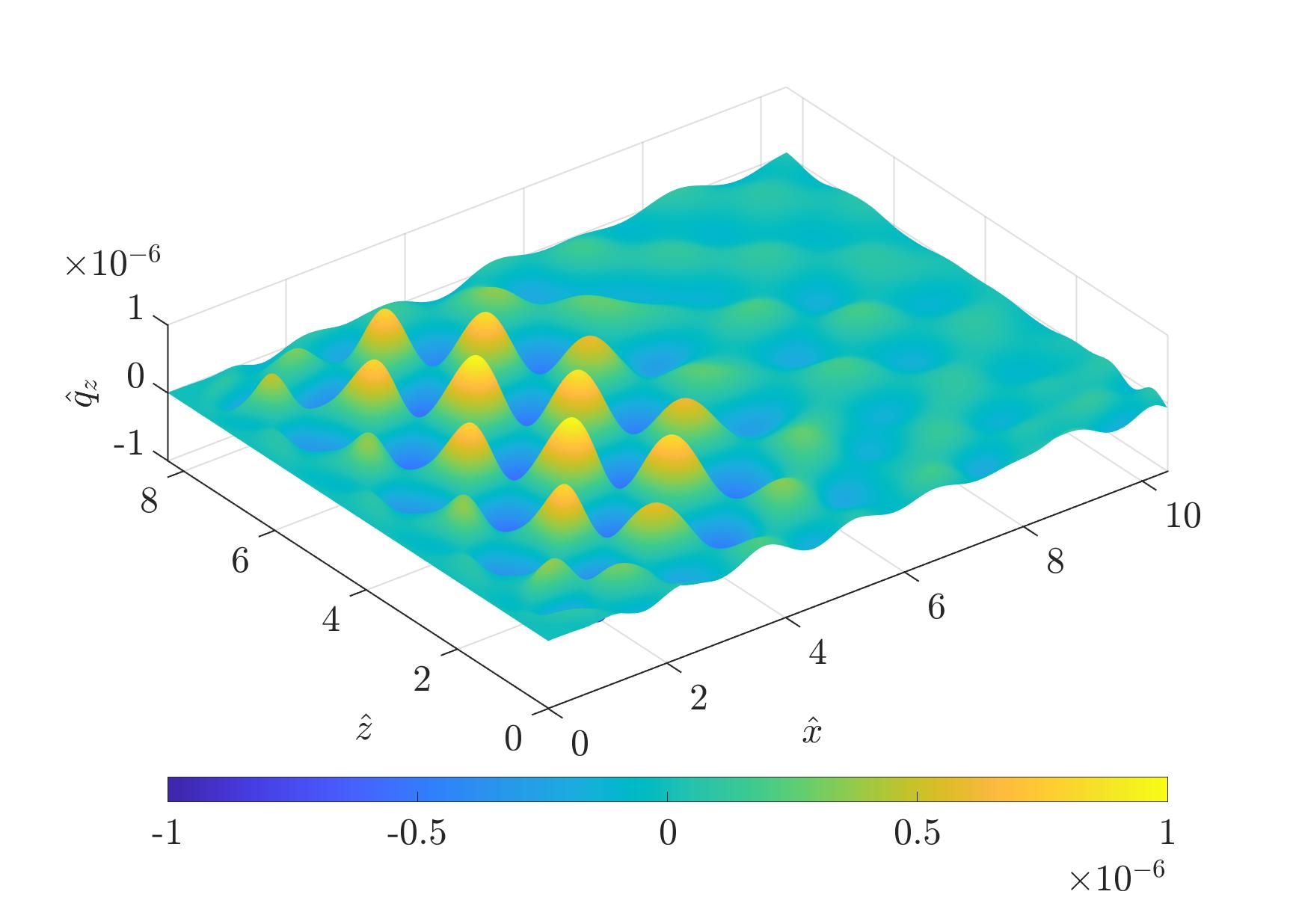}
    \caption{\label{fig:qz_kx3_kz3_ST} Spatial evolution of the span-wise flow rate $\hat{q}_z$ for the base case ($\hat{k}_x = 3$, $\hat{k}_z = 3$) when surface tension is activated \textcolor{black}{in the IBL solver. The same computation in absence of surface tension leads to $\hat{q}_z = 0$ everywhere, and therefore it is not included here.}}
\end{figure}

For example, it is interesting to note that the model preserves 2D solutions if the boundary conditions sets $\hat{q}_z |_{x=0} = 0$ as in the base state: in this case, all flux and source terms in \eqref{eq:BLEW_Mz} cancel resulting in $\partial_{\hat{t}} \hat{q}_{\hat{z}}=0$ and hence no span-wise flow can be produced. The DNS confirms this is not an artifact of the model as the transverse velocity component, $w$, also remains null downstream. Similarly, it can be shown that when the surface tension terms (third derivatives in \eqref{eq:BLEW_Mx} and \eqref{eq:BLEW_Mz}) become dominant as in the case shown in Fig.~\ref{fig:LSA_vs_BLEW_vs_OpenFOAM}, the system responds with a variation of the stream and span-wise flow rates. This is illustrated in Fig.~\ref{fig:qz_kx3_kz3_ST} with the span-wise flow rate $\hat{q}_z$ distribution for \textcolor{black}{case 2} ($\hat{k}_x = \hat{k}_z = 3$). The 3D capillary damping is therefore linked to \textcolor{black}{a} production of span-wise flow rate, which simultaneously enhances the non-linear interactions, further damping the waves. In this regard, it is remarkable that the IBL model captures the interaction between stream-wise and span-wise components of the flow in agreement with the Navier-Stokes equations.

\textcolor{black}{These results are also in qualitative agreement with the predictions of the LSA in Fig.~\ref{fig:h02_delta150}. The trend is correctly captured since an increase of the span-wise wave number $\hat{k}_z$ is beneficial for the stability of the system in all frameworks.} On the other hand, the exact stability regions have not been correctly identified: for instance, cases 1 and 2 --unstable according to LSA-- resulted to be stable in the non-linear framework (the waves decay as they travel downstream with both the IBL and DNS solvers). The mismatch between linear and non-linear tools is not surprising and can be attributed to various causes. First, the LSA neglects the non-linear terms in the governing equations, which according to previous studies\cite{Ivanova2022_BLEW3D} might have an stabilization effect on the flow. Second, the LSA relies on infinitesimal perturbations on $(\hat{h}, \hat{q}_x , \hat{q}_z)$ and the \emph{initial} response of the flow, while the IBL and DNS models impose finite perturbations ($A = 0.05$) and consider the film response over larger time scales. Third, it is well known that IBL model fails to predicting the location of the neutral stability curves because high order corrections of the velocity profile are not considered\cite{Ruyer-Quil1998}. Nevertheless, the presented results validate the IBL approach in conditions relevant to coating process and shows that it can be useful to characterize the evolution of 2D and 3D perturbations in industrially relevant conditions.

\begin{figure}
    \centering
	\includegraphics[width=\linewidth]{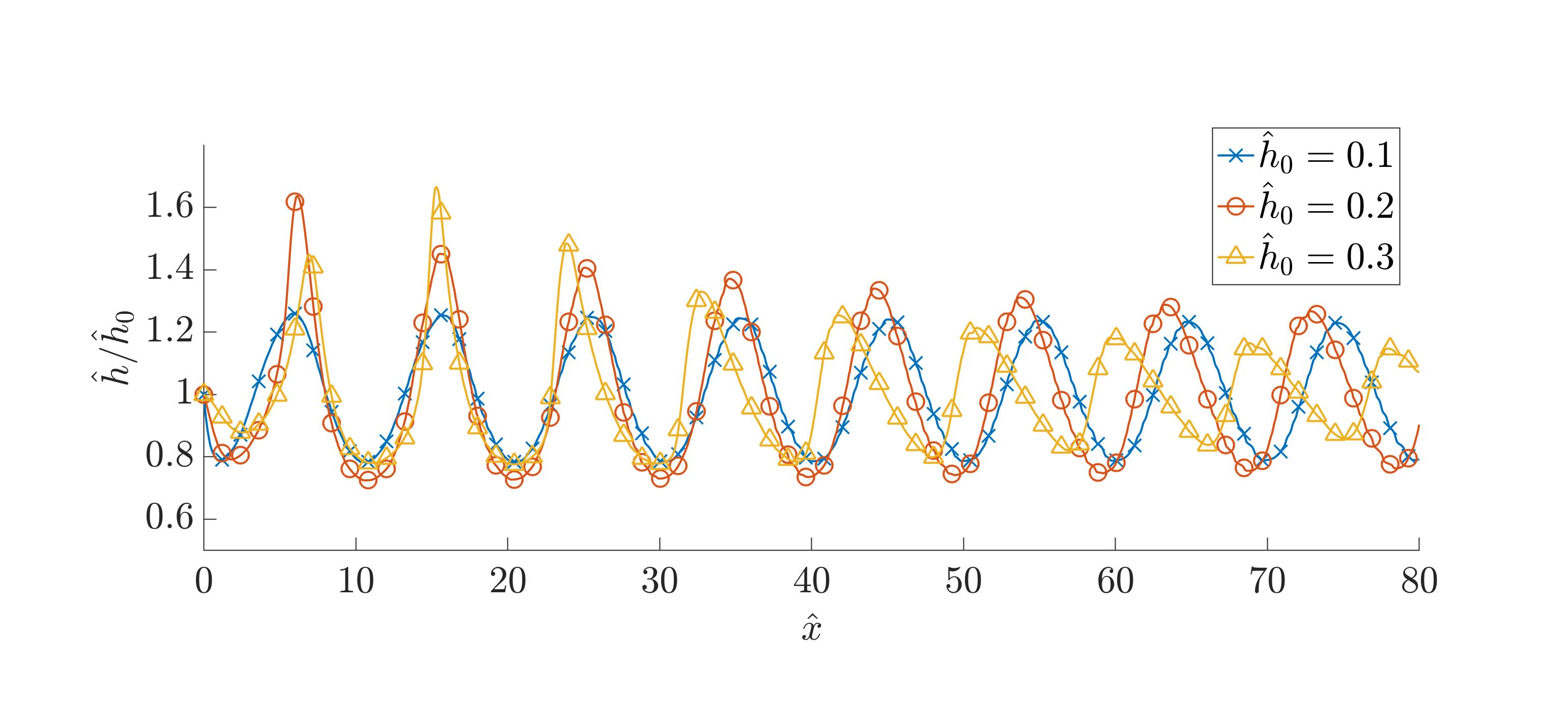} 
	\caption{\label{fig:Case1_OF} Stream-wise evolution of three liquid films characterized by $\delta = 153$ and with a dimensionless frequency of $\hat{f}=0.1$ and amplitude of $A = 0.2$ as in Ivanova et al.\cite{Ivanova2022_BLEW3D}, except for $\hat{h}_0$=0.3 in which the amplitude has been fixed at 0.1 to avoid breakup of the wave crest. The vertical axis has been normalized with the nominal film thickness $\hat{h } / \hat{h}_0 $.}
\end{figure}

Finally, we close with a note on the non-linear \textcolor{black}{stabilizing mechanism} documented in Ivanova et al\cite{Ivanova2022_BLEW3D} using the same IBL model in absence of surface tension. The analysis of this non-linear damping mechanism is carried out on the basis of DNS computations \textcolor{black}{using water and air as working fluids}. Fig.~\ref{fig:Case1_OF} depicts the film thickness profiles normalized with their respective nominal value $\hat{h}_0$ for three different configurations \textcolor{black}{characterized by} $\hat{h}_0$=[0.1;0.2;0.3]. \textcolor{black}{The substrate velocity is $U_p = 1.46$ m/s, leading to a reduced Reynolds number $\delta = 153$}. The stream-wise flow rate is perturbed at a dimensionless frequency $\hat{f}=0.1$ and relative amplitude $A=0.2$, so that the flow remains strictly bi-dimensional. 

The non-linear damping previously reported for the same conditions\cite{Ivanova2022_BLEW3D} is well captured in our DNS computations, and its effect increases with the film thickness $\hat{h}_0$, in agreement with the previous findings. The non-linearities are enhanced by the initial growth of the perturbation, which on the other hand increases with the nominal film thickness $\hat{h}_0$. This stabilizing mechanism is mostly driven by the velocity difference in the bulk of the film (approximately the velocity of the moving plate) and the wave crest (slower due to the action of gravity acting towards the left in this figure). The wave shape is also significantly altered due to gravity, which deflects it downwards as it is slowed down. In the extreme case of large waves, part of the crest \textcolor{black}{might detach} from the bulk flow and break up in droplets. The IBL solver cannot cope with this type of flows, and therefore needs to be carefully applied in these flow regimes.

\section{Conclusions}
\label{sec:conclusions}

We have investigated the evolution of three-dimensional disturbances on liquid films dragged by upward-moving substrates, a fundamental flow configuration in coating processes such as the air-knife and slot-die coating. We analyzed the downstream evolution of 2D and 3D disturbances using two-phase Direct Numerical Simulations (DNS) and the Integral Boundary Layer (IBL) thin film model. For the second, we consider both Linear Stability Analysis (LSA) and nonlinear analysis by simulating the response of the full model to finite perturbations.

The sensitivity of the IBL and DNS solvers to the numerical configuration has been examined in detail. The agreement between these has been found to be excellent considering the disparity in complexity and computational cost \textcolor{black}{and considering that the investigated conditions are well outside the range of validity for the asymptotic expansion underpinning the IBL derivation}. The IBL solver has been validated with three-dimensional non-periodic waves against DNS computations carried out with OpenFOAM, and the implemented pseudo-spectral algorithm to compute the capillary terms has proved to effectively constrain the numerical dispersion without compromising the accuracy of the computations.  

The linear stability analysis of the IBL governing equations allowed for deriving the amplification contours $\hat{\omega}(k_x,k_z)$ for different film thickness $\hat{h}_0$ and reduced Reynolds numbers $\delta$. The stability of the system with respect to the stream and span-wise wave numbers has been analyzed within a typical range of operational conditions of coating processes. 
The results have been confronted with the predictions of the nonlinear IBL and DNS solvers for three cases to investigate the role of capillary damping. The span-wise disturbances turned out to have a stabilizing effect on the film due to the generation of a span-wise flow rate that drives the liquid from crests toward troughs. 

\textcolor{black}{The LSA correctly predicted the stabilizing effect of higher span-wise wave numbers, although the stability regions are not in agreement with the nonlinear analysis described in Section~\ref{sec:results_nonlinear}.} Finally, the nonlinear damping mechanism reported in Ivanova et al.\cite{Ivanova2022_BLEW3D} has been confirmed with our DNS computations. However, it is found to be weakly active within the operating window under study. This effect is more pronounced for larger film thicknesses and larger amplitudes.

The results obtained with the synthetic cases in this work can be extrapolated to \textcolor{black}{different industrial coating processes. T}he validated nonlinear film model proves to be a helpful tool to investigate the dynamics of \textcolor{black}{paints and coatings} while still in liquid phase after deposition, especially for non-periodic three-dimensional \textcolor{black}{defects}. The results even suggest the possibility of generating intentional 3D disturbance patterns in coatings to promote the smoothing of the free surface by capillarity\textcolor{black}{, and open the door for using the model to train active control strategies as in Cimpeanu et al.\cite{Cimpeanu2021}.} We are currently working on more complex configurations involving interfacial shear stress and pressure gradient imposed by an external gas flow, as it occurs in the jet wiping process.

\begin{acknowledgments}
    D.Barreiro-Villaverde is financially supported by Xunta de Galicia with the pre-doctoral grant "Programa de axudas á etapa predoutoral" (ED481A-2020/018) and the research project is founded by Arcelor-Mittal. The authors also wish to thank the “Red Española de Supercomputación” for the attribution of special computational resources at FinisTerrae III (CESGA) (IM-2021-3-0012). 
\end{acknowledgments}

\section*{Data availability}
     The data that support the findings of this study are available on request from the corresponding author. The data are not publicly available due to privacy restrictions.

%\nocite{*}
%\section*{References}

%merlin.mbs aipnum4-1.bst 2010-07-25 4.21a (PWD, AO, DPC) hacked
%Control: key (0)
%Control: author (8) initials jnrlst
%Control: editor formatted (1) identically to author
%Control: production of article title (0) allowed
%Control: page (1) range
%Control: year (1) truncated
%Control: production of eprint (0) enabled
%

\bibliography{}% Produces the bibliography via BibTeX.

\end{document}